**Extensive characterization of a high Reynolds number decelerating boundary layer using advanced optical metrology.**

**Authors:** C. Cuvier[1,7], S. Srinath[1,6], M. Stanislas[1,6], J. M. Foucaut[1,6], J. P. Laval[1,7], C. J. Kähler[2], R. Hain[2], S. Scharnowski[2], A. Schröder[3], R. Geisler[3], J. Agocs[3], A. Röse[3], C. Willert[4], J. Klinner[4], O. Amili[5], C. Atkinson[5], J. Soria[5].

[1] Université Lille Nord de France, FRE 3723, LML-Laboratoire de Mécanique de Lille, F-59000 Lille, France,
[2] Bundeswehr University Munich, Institute of Fluid Mechanics and Aerodynamics, Neubiberg, Germany,
[3] German Aerospace Centre (DLR), Institute of Aerodynamics and Flow Technology, Göttingen, Germany,
[4] German Aerospace Center (DLR), Institute of Propulsion Technology, Köln, Germany,
[5] Monash University, Australia,
[6] Centrale Lille, F-59650 Villeneuve d'Ascq, France
[7] CNRS, FRE 3723 -LML- Laboratoire de Mécanique de Lille, F-59650 Villeneuve d'Ascq, France.

1. Introduction

Over the last years, the observation of large scale structures in turbulent boundary layer flows has stimulated intense experimental and numerical investigations with the aim of characterizing not only the topological features of the coherent structures but also their dynamics. Nevertheless, our understanding of turbulence near walls, especially in decelerating situations, is still quite limited. This is partly due to the lack of comprehensive experimental data at sufficiently high Reynolds number. Very often, the facilities are too small to reach high Reynolds numbers and to let the boundary layer develop enough to reach some state of "equilibrium" where theoretical approaches can be relevant. Moreover, measurements are generally quite limited leading to a lack of detailed characterisation of the flow itself but also of the boundary conditions, which makes the data very difficult to use in practice, both for physical understanding and for models validation.

Adverse pressure gradient boundary layers are particularly problematical. There are indeed some careful experiments available in the literature on APG wall flows (e.g., (Ludwig & Tillman, 1950), (Perry & Shofield, 1973), (Skare & Krogstad, 1994), (Elsbery, et al., 2000), (Perry, et al., 2002), (Nagib, et al., 2004), (Maciel, et al., 2006), (Aubertine, 2006), (Rehgozar et al., 2011), (Joshi et al., 2014) (Knopp, et al., 2015)), but most of them provide only single-point turbulence statistics at a few stations, and generally only for one or at best two velocity components. In fact there is virtually no information on the turbulence integral scales which appear in even some of the simplest turbulence models. However, there have recently been several important breakthroughs which have dramatically transformed our view of APG Turbulent Boundary Layers (APG-TBL). The first was the confirmation that while the peak in turbulence energy production is maximum around y+ = 10-20 in ZPG Turbulent Boundary Layers (ZPG-TBL), in APG, the production peak moves away from the wall. Second, this outward displacement of the turbulence intensity peak is usually correlated with the downstream evolution of an inflexional mean velocity profile, ( (George, et al., 2010a), (George, et al., 2010b), Figure 1). The location of this inflection usually follows the peak in turbulence intensity. It appears to have been previously noticed only by (Elsbery, et al., 2000). But it seems to be characteristic of all APG-TBL's if allowed to evolve for a large enough distance (which explains why it has not been previously noticed in the typical low Reynolds number short wind tunnel experiments). Note also that it seems to be quite independent of whether the boundary layer ultimately separates. The existence of any inflectional profile in fluid mechanics is a well-known game changer, since very different instability dynamics come into play. Exactly how this happens in APG-TBL is not understood yet, but understanding it is of major practical importance.

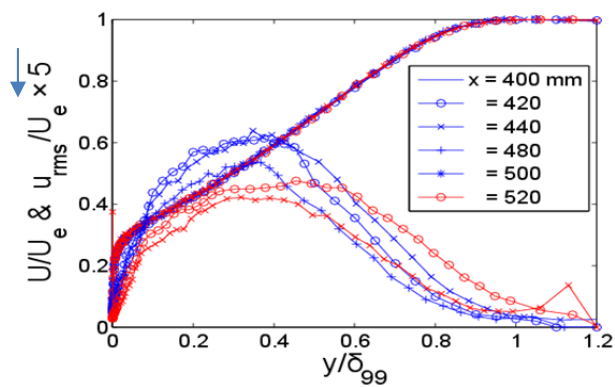

**Figure 1 : Results of (Skare & Krogstad, 1994) scaled by (George, et al., 2010a) showing the inflection of the mean velocity profile and the turbulence peak.**

The third new phenomenon came from the DNS performed during WALLTURB (Marquillie, et al., 2008) which provides a clue to solve the riddle of its origins. Using this DNS and stability theory, they were able to establish a sudden growth of turbulence intensity arising from inflectional instabilities of the streaks at the imposition of the APG (Marquillie, et al., 2011). It seems quite likely that this phenomenon has some generality. Exploring the relations among these new phenomena is important from the modelling point of view.

A great deal of work has been done on boundary layers in the past 50 years, but very few at high Reynolds numbers. For example, most experiments have been performed at values of $Re_\theta$ < 10,000. This limit on the Reynolds number of experiments means there is quite likely a critical gap in our knowledge, since recent theoretical arguments suggest strongly that it is only above this value that the true inertial nature of the high Reynolds number boundary layer begins to be manifested ( (George & Castillo, 1997), (Tutkun, et al., 2010)). Measurements beyond this range have only very recently been able to successfully resolve the near wall region in ZPG boundary layers (say y+ < 100, e.g., (Österlund, 1999), (Carlier & Stanislas, 2005), (De Silva, et al., 2014) and even most of these new measurements are under debate.

Computer simulations (DNS) have also been restricted to quite modest Reynolds numbers, limiting the universality of their findings. Happily, recent simulations in channel flow from Lee and Moser (2015), at $Re_\tau \approx 5{,}200$ are beginning to remove this limitation. There has been considerable effort over the past four decades to apply a deterministic approach to near wall turbulence (Jeong, et al., 1997), (Jiménez & Pinelli, 1999), (Schoppa & Hussain, 2002), (Kawahara, et al., 2003), (Jimenez & Moin, 1991), (Hamilton, et al., 1995), (Jiménez & Simens, 2001), (Waleffe, 2001), (Kawahara & Kida, 2001) among others). In spite of this, our knowledge of what happens in an adverse pressure gradient boundary layer (especially approaching separation) is far less complete.

Finally, there is much less agreement on how the near-wall region interacts with the outer flow, although it is clear that in attached boundary layers there is a net transport of energy away from the wall (Jiménez & Pinelli, 1999). In addition, the near wall may contribute to the formation of large-scale outer structures in the form of hairpin packets (Adrian, 2007). It is also known that the outer flow modulates the inner layer ( (Tutkun, et al., 2010), (Marusic, et al., 2010)), forcing some of its characteristics to scale purely in outer units.

In this type of flow, due to the large length of the structures (approximately 7-14$\delta$) it is very difficult to analyse them reliably since both a large field of view and a high spatial resolution are simultaneously required to measure all relevant spatial scales. Moreover, a low uncertainty of the experimental techniques is required as the disturbance of the velocity field can be quite weak. As a consequence of the recent technological progress, the required measurement uncertainty can be reached with sophisticated image analysis techniques but the resolution of the structures requires a relatively large number of cameras which is not available at any one research organisation alone.

The aim of the present project was to combine the equipment and skills of several teams to perform a detailed characterization of a large scale turbulent boundary layer with a spatial resolution not reachable with the equipment of one single team. The fundamental objective was to resolve and characterize the turbulence structures in an adverse pressure gradient turbulent boundary layer flow. A second aim of the project was to record well converged statistics and to characterize boundary conditions in order to use the experiment as a test case for CFD. In addition to two-component two-dimensional PIV measurements in a streamwise/wall-normal plane with 16 sCMOS cameras, stereoscopic PIV measurements in a spanwise/wall-normal plane were performed at two specific locations in order to resolve the spanwise velocity distribution in this APG-TBL flow. To complement these large field of view measurements, time resolved near wall velocity profiles were obtained in order to determine the wall-shear stress and its fluctuations at some specific locations along the wall.

Due to the richness of the resulting dataset, it is out of reach to present all of it in a single paper. The aim of the present contribution is consequently to provide the main results of this unique experiment with an emphasis on the statistics and the boundary conditions. The analysis of the turbulence organization will be the subject of a following paper. The objective is in fact to provide modellers with a carefully characterised flow which will allow them either to validate their RANS model on a challenging test case or to be able to compare their LES predictions of the flow organisation with the experiment. For that purpose, pressure distributions on both lower and upper wall were measured, detailed upstream boundary conditions are provided, the 2D character of the flow has been checked by performing measurements near the side walls and complementary measurements have been performed in the streamwise plane of symmetry, upstream of the APG in order to provide the modellers with as a complete picture of the flow as possible. All these results are described in quite detail in the following to try to avoid any ambiguity on the experimental conditions and on the accuracy of the results. The experimental facility is described first in section 2. Then the details on the different PIV set-ups used are grouped in section 3. For the purpose of conciseness, the experiments performed to characterise the boundary conditions and the side wall flows are reported in annex in section 9. Section 4 is devoted to the pressure distribution in the test section, section 5 to the flow on the FPG wall, section 6 to the flow on the APG wall, section 7 gives synthesis plots of the results before conclusions are given in the last section. Finally section 8 provides some conclusions.

## 2. The EuHIT experiment

In the Framework of the EuHIT European project (www.euhit.org), experiments were performed in the LML boundary layer wind tunnel which has a test section length of 20.6 m. Figure 2 and Figure 3 show respectively a front and a top view of the LML turbulent boundary layers (TBL) wind tunnel (Carlier & Stanislas, 2005). The test section is slightly longer than 21 m and is now, thanks to specific CISIT/CNRS funding, transparent on all sides with high quality 10 mm glass providing full optical access (Figure 4). The cross-section is 2 m wide and 1 m high with a free-stream velocity range from 1 to 9.4 m/s measured 100 mm downstream of the tunnel test section entrance. The wind tunnel can be used in a closed loop configuration with temperature regulation or opened to the outside. In the first configuration, the free-stream velocity is regulated within ±0.5 % and the temperature within ±0.15 °C. The boundary layer is tripped at the tunnel entrance using a 4 mm spanwise cylinder fixed with silicon on the bottom wall followed by 93 mm of Grit 40 sandpaper (mean roughness 425 μm). The boundary layer that develops on the top wall is also tripped but only with 93 mm of Grit 40 sandpaper. All the glasses and surfaces at the bottom wall are adjusted so that there is no step bigger than 0.1 mm (corresponding to $2^+$ at the maximum velocity). The top and bottom walls were also adjusted so that they are perfectly parallel (locally less than ±0.1°). The same is true for the lateral walls.

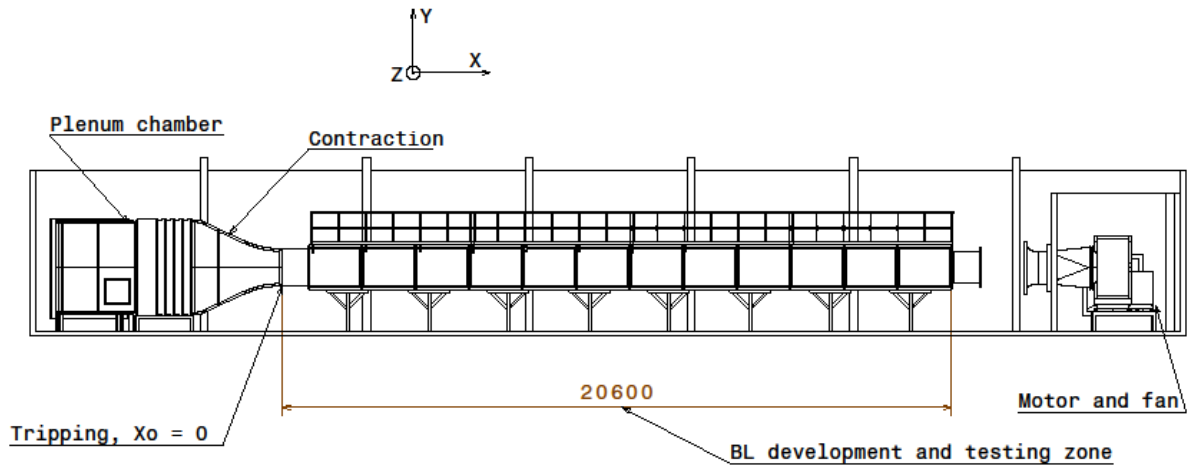

**Figure 2: Sketch of the LML wind tunnel in front view.**

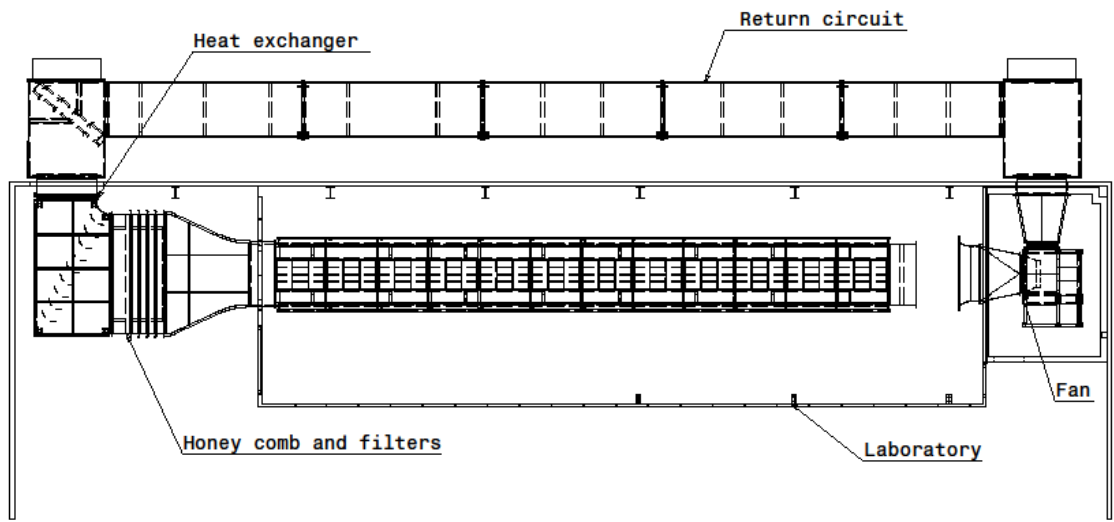

**Figure 3: Sketch of the LML wind tunnel in top view.**

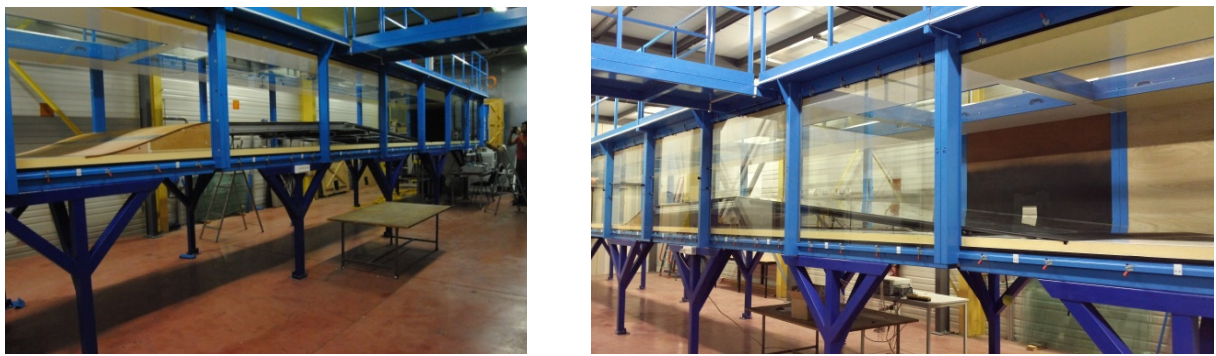

**Figure 4: Test section of the LML wind tunnel with EuHIT model mounted on the floor.**

For the EuHIT test campaign, the wind tunnel was used in the closed loop configuration and the bottom wall was equipped with a specifically designed ramp model approximately 7 m long. The model used is the LML-AVERT ramp (Cuvier, et al., 2014) on which the final flap was replaced by a 3.5 m long plate (see Figure 5).

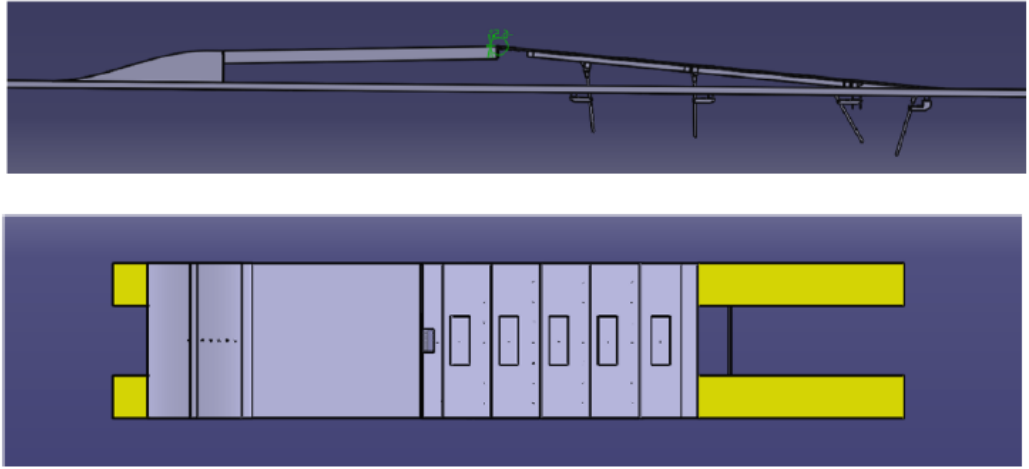

**Figure 5: a) Profile view and b) Top view of the modified LML-AVERT platform installed in the wind tunnel, showing the position of the windows.**

Figure 6 gives a quantitative sketch of the model. The leading edge of the ramp was placed 9.4 m downstream from the beginning of the test section. The converging part has a contraction ratio of 0.75. It is followed by a 2.14 m long flat plate at 1.5° angle with respect to wind tunnel floor. The plate is composed of four interchangeable parts with a fairing accuracy smaller than 0.1 mm (less than $2^+$ at maximum velocity) and a flatness accuracy below ±0.1° on 200 mm. The last 3.5 m flat plate is set at -5° compared to the wind tunnel floor. It is composed of 7 plates with the first one 240 mm long fabricated in aluminium, the following five are fabricated from plexiglass (four of 625 mm and one of 515 mm) and the last one out of aluminium (210 mm long and 2 mm thick). All the adjustments between plates are better than 0.1 mm and the surface quality of this 3.5 m plate is below ±0.1° on 200 mm. To minimize leaks, a black aluminium tape 0.05 mm thick was set spanwise on all the junctions between two plates. Also to minimize the vortex that develops on the side walls, a gasket is fixed between the side walls and the ramp model. Each plexiglass plate is equipped with a 240 by 625 mm² insert to allow specific near wall measurements during the campaign. The ramp is also equipped with pressure taps: 27 for the streamwise pressure distribution and 24 for four transverse pressure distribution stations, two on the 1.5° plate and two on the -5° plate.

The exact ramp coordinates in the (X, Y, Z) global wind tunnel frame are given below (X = 0 corresponding to wind tunnel entrance). $X_r$ is defined as $X_r = X - X_{le}$, with $X_{le}$ the ramp leading edge position ($X_{le}$ = 9400 mm). The converging part is defined by equation (1) with $X_r$ and Y in mm.

$$\begin{cases} Y = -\dfrac{500}{1200^3} X_r^3 + \dfrac{750}{1200^2} X_r^2 & \text{for } 0 \le X_r \le 1200\,mm \\ Y = 250\,mm & \text{for } 1200 \le X_r \le 1330\,mm \end{cases} \qquad (1)$$

At the end of the converging part, the articulation has a radius of 10 mm which can be modelled by a sharp corner. This articulation is followed by a flat plate at +1.5° from the floor on a length of 2140 mm. At the end of this flat plate, the articulation corresponds to a circular arc of radius 10.25 mm with the centre at $X_r$ = 3470 mm, Y = 295.77 mm and a -6.5° rotation to join the 3486 mm flat plate which follows at −5° from floor. At the end of this plate, the connection with the wind tunnel floor is done with a curvature radius of about 7.27 mm centered at $X_r$ = 6949.04 mm, Y = 5.14 mm.

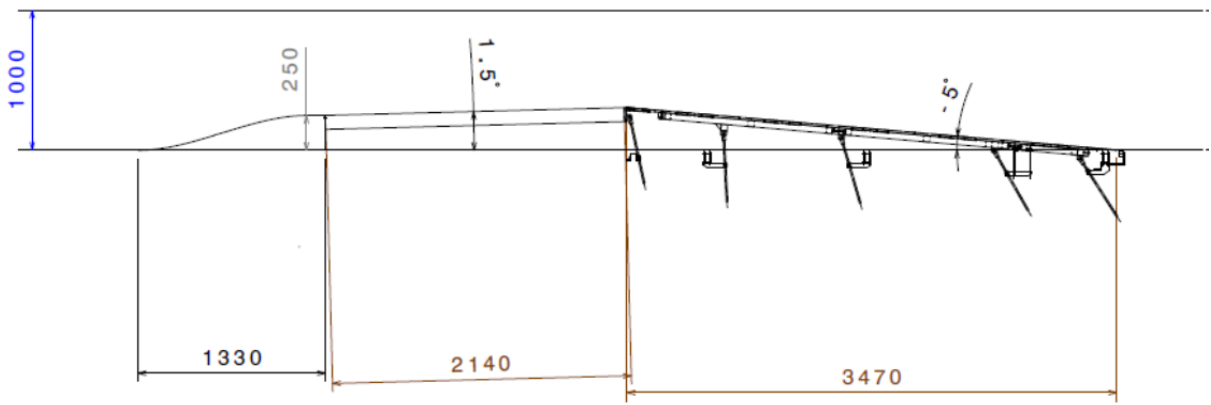

Figure 6: Sketch of the ramp model showing the main dimensions

Table 1 gives the coordinates of the ramp (including the curvilinear coordinate "s", with leading edge as origin) at some representative points between the leading and trailing edges. Also, the full shape is available as a CAD file in the file "Files/ramp_euhit.stp" at https://turbase.cineca.it/turbase/default/#/view_dataset/43 .

| Xr (mm) | Yr (mm) | s (mm) | Xr (mm) | Yr (mm) | s (mm) |
|---:|---:|---:|---:|---:|---:|
| 0.00 | 0.00 | 0.00 | 1330.00 | 250.00 | 1360.69 |
| 100.00 | 4.92 | 100.16 | 3470.00 | 306.02 | 3500.69 |
| 200.00 | 18.52 | 201.10 | 3470.27 | 306.02 | 3500.96 |
| 300.00 | 39.06 | 303.21 | 3470.54 | 306.00 | 3501.23 |
| 400.00 | 64.81 | 406.48 | 3470.80 | 305.99 | 3501.50 |
| 500.00 | 94.04 | 510.66 | 3471.07 | 305.96 | 3501.76 |
| 600.00 | 125.00 | 615.35 | 3471.16 | 305.95 | 3501.85 |
| 700.00 | 155.96 | 720.03 | 6943.90 | 2.13 | 6987.85 |
| 800.00 | 185.19 | 824.21 | 6944.36 | 1.70 | 6988.49 |
| 900.00 | 210.94 | 927.48 | 6945.40 | 0.97 | 6989.76 |
| 1000.00 | 231.48 | 1029.59 | 6946.55 | 0.44 | 6991.03 |
| 1100.00 | 245.08 | 1130.53 | 6947.77 | 0.11 | 6992.30 |
| 1200.00 | 250.00 | 1230.69 | 6949.04 | 0.00 | 6993.56 |

**Table 1: Coordinates of the ramp from its leading to trailing edges. The curvilinear coordinate s along the ramp surface is also given for completeness.**

## 3. PIV set-ups

In order to carefully characterize the flow, several complementary PIV experiments were performed. The most challenging one: the large field 2D2C PIV measurement in the APG region will be detailed here. The other set-ups will be summarized as they either are standard or they are described in detail in other publications. The set-ups used to characterize the boundary conditions (upstream BC and corner flow) are described in the annex in section 9. For all the experiments, seeding was provided globally in the closed circuit wind tunnel. It consisted of an evaporated-recondensed water-glycol mixture and was introduced in the diffuser downstream of the 20 m long test section just upstream of the fan. The size of the aerosol droplets was estimated at 1μm with a lifetime in the order of 10 minutes. Table 2 summarizes the main parameters used for PIV recording in the different experiments.

### 3.1 Large field streamwise 2D2C PIV (LFStW PIV).

The large field 2D2C PIV measurements were performed in a streamwise vertical plane of 3.5 m long placed in the middle of the test section (XY plane at Z=0). A specific illumination set-up was designed for that purpose which is sketched in Figure 7. The LML BMI YAG laser system was installed on the side of the wind tunnel, at 11.6 m from the inlet of the test section. Two of the four laser cavities where used to generate two pulses of 200 mJ each. A set of mirrors and lenses allowed the generation of a light sheet under the wind tunnel in the vertical plane of symmetry. This light sheet was introduced inside the tunnel via a mirror at 45° placed downstream of the ramp model and reflecting upward. A slot slightly larger than the light sheet and closed by a hermetic box below the tunnel floor allowed the light to enter the wind tunnel. The slot was 5 mm wide and started just downstream of the end of the ramp. The second mirror was inside the test section, 87 cm downstream of the end of the ramp. It was 20 mm wide, 450 mm long and only 250 mm of it were above the tunnel floor. The backside of the mirror was equipped with an aerodynamic profile to limit the vibration due to vortex shedding. This set-up provided a light sheet of 300 mm in height at the beginning of the field of view and 1 mm in thickness all along the 3.5 m of the APG region. The light sheet was carefully set tangent to the ramp wall to minimize light reflection from it. With this set-up, the whole field of view of the 16 SCMOS camera used in the experiment (3500 mm x 250 mm) could be covered with enough illumination.

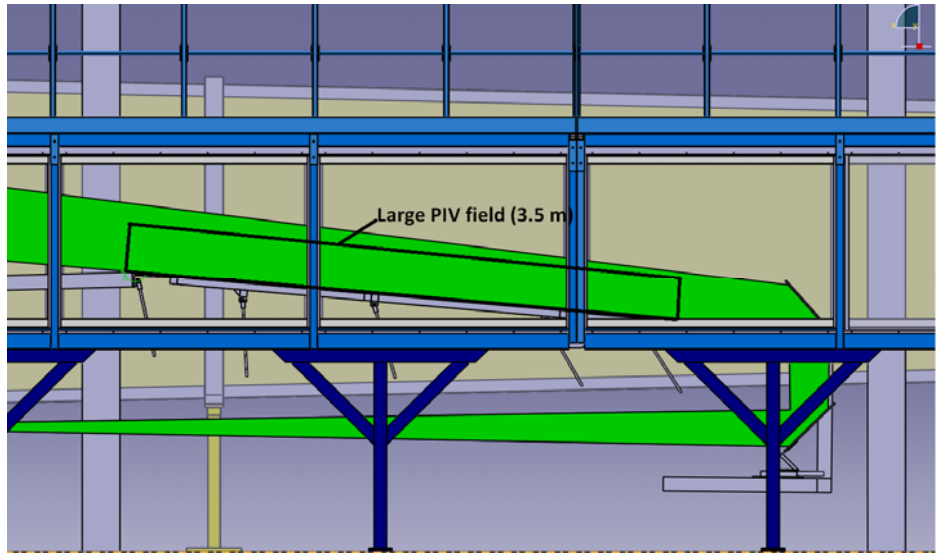

**Figure 7: CAD model of the light sheet of the large APG field of view**

The length of the measurement domain of the multi-camera 2D2C-PIV is highlighted by the black rectangle in Figure 7. It is 3.5 m long and 25 cm high to ensure the possibility of capturing very large scale turbulent structures with lengths of more than 10 boundary layer thickness $\delta$. Figure 8 gives an overall view of the model in the wind tunnel with the 16 sCMOS cameras mounted on the side of the tunnel. All the cameras were fixed on the same 3.5 m long X95 bench and mounted on Manfrotto 410 articulations to allow the tuning of the cameras normal to the side glass wall of the tunnel and parallel to the ramp surface. The cameras were mounted at 90° so that the 2560 pixels side of the sensor was aligned with the wall-normal direction to obtain the best resolution on the 250 mm of height. The field of view of each camera was 230 mm along the wall and 273 mm in the wall normal direction. A common region of about 10 to 20 mm was set between each camera to obtain a continuous field. All cameras except three were equipped with macro planar 100 mm Zeiss lens at a working distance of 1680 mm. To avoid the shadow of the wind tunnel side pillars (see Figure 8), cameras number 6 (camera 1 being the most upstream one) and 13 were equipped with 85 mm Zeiss lens at a working distance of 1445 mm and camera 14 with a 50 mm Zeiss lens at a working distance of 1050 mm. The f# number was 4 for all cameras except for cameras 6, 15 and 16 where it was fixed at 2.8, 4.8 and 4.8 respectively.

A total of 30 000 2D2C velocity samples were recorded at a frequency of 4 Hz for the two free stream velocities of 5 and 9 m/s, representing 20 To of raw images.

The PIV images were processed with a modified version of the Matpiv toolbox at LML. First the mean background images were mapped (basic pinhole model) and the reflection (wall position) was manually fitted with a line. A mesh was then build above this line in the mapped images (spacing of about 1.07 by 1.07 mm corresponding to 10 by 10 pixels) and projected on each camera with the pinhole models. The analysis was then done with these projected grids. The cross-correlation PIV analysis ( (Willert & Gharib, 1991), (Soria, 1996))was done with four passes (one 64x64, two of 32x32 and a final one of 24x24). Before the final pass, image deformation ( (Scarano, 2002), (Lecordier & Trinité, 2004)) was used with a cubic b-spline interpolation of the grey level and bilinear interpolation for the displacement to improve the result quality. Also background division (Raffel, et al., 2007) was used to limit the effect of the laser reflection and the camera noise. The final field has then 3250 vectors along the wall and 238 in the wall normal direction.

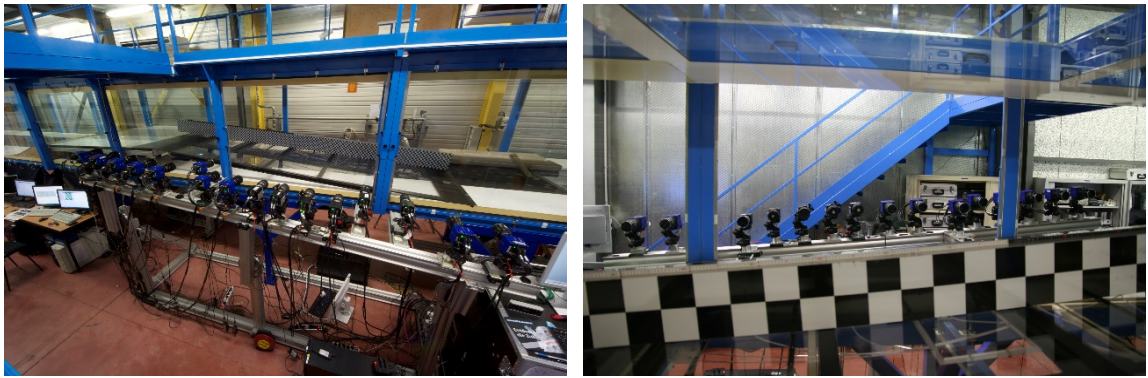

**Figure 8: Set of 16 SCMOS cameras mounted on the side of the wind tunnel. The target used for calibration is visible in the right image.**

### 3.13.2   Spanwise SPIV (SpW SPIV).

In order to characterize the spanwise structure of the flow in the APG region, Stereo PIV measurements with two SPIV systems side by side to increase the field of view were performed in a transverse plane (normal to the wall and to the flow) at two stations along the diverging ramp. In this section, the local reference frame is used, with *x*-axis parallel to the model wall with the origin at the beginning of the APG plate (corresponding to s = 3500 mm).

The global field-of-view (FOV) was z ≡ [-162; 162] mm in span (z = 0 mm being the centerline of the test section) and y ≡ [0;120] mm in the wall normal direction. For both SPIV measurement positions, four sCMOS cameras with 2560 x 2160 px resolution and 16 bit dynamic range were used. They were equipped with *Zeiss* lenses with a focal length of f = 180 mm and an aperture number of $f_{\#}$ = 11 and were working in forward scattering mode. All had direct-to-disk data acquisition at 5 Hz image rate.

For the illumination of the tracer particles, a double-PIV Nd:YAG laser system from *BigSky* with four laser cavities in total was used. This laser system enables the possibility to combine two oscillators each and generate a laser pulse of up to 400 mJ with a pulse length of 9 nsec. Proper light-sheet optics combining a telescope and a cylindrical lens were used in order to form a light sheet plane with an almost constant thickness of ~2 mm in the whole FOV. Figure 9 shows a photo of the laser light-sheet and the stereo camera set-up at the second (downstream) position at the LML wind tunnel.

The first measurement plane position was located at x = 595 mm downstream of the edge of the inclined flat plate (corresponding to s= 4095 mm). The second measurement plane position was located further downstream at x = 2192 mm from the articulation (s = 5692 mm).

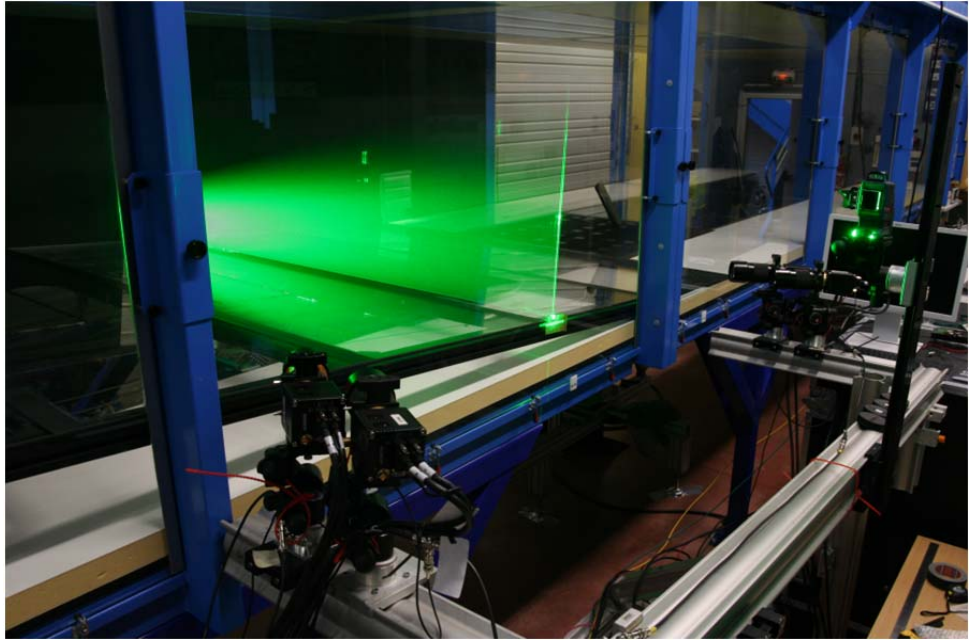

**Figure 9: Stereo PIV set-up at the second downstream position (x = 2192 mm) with two pairs of sCMOS cameras on Scheimpflug mounts and laser sheet normal to the wall and mean flow directions.**

For both streamwise positions, two Reynolds numbers were measured. Therefore the free-stream velocity $U_\infty$ was again: 5 m/s and 9 m/s. The temperature was 20 °C for all cases. A total number of more than 30,000 2D3C velocity snapshots (each in 6 blocks of 5040 images) were recorded for each Reynolds number and position of the combined SPIV camera set-up. Calibration and mapping of the two FOV were realized using a glass target with a squared regular grid of 10 mm x 10 mm.

The acquired particle images were analyzed using the *PIVview3.60* software, employing iterative multi-grid interrogation ( (Willert & Gharib, 1991), (Soria, 1996)) with image deformation (Scarano, 2002). A final window size of 24 x 24 $px^2$ (1.33x1.33 $mm^2$) at a step-size of 6 px (0.33 mm, i.e. 75% window overlap) were used for this analysis. This results in ~390,000 instantaneous three-component velocity vectors per snapshot in the combined FOV of the two SPIV systems.

### 3.3 Time-resolved high magnification PIV (TRHM PIV).

In order to characterize more accurately the very near wall region in the APG-TBL, high magnification/high repetition 2D2C PIV experiments were conducted at a few stations along the diverging ramp. As illustrated in Figure 10, boundary layer profile measurements were also obtained along the tunnel centerline (Z=0) at two positions upstream of the model (at a distance from the tripping device of 3.2 m and 6.8 m respectively), labelled (A) and (B), and at three positions in the APG region at the center of the inserts along the ramp, labelled (1), (3) and (4). The positions of the profiles in the APG region are located respectively at 0.483 m, 1.733 m and 2.358 m from the beginning of the APG region corresponding to s= 3.983, 5.233 and 5.858 m respectively.

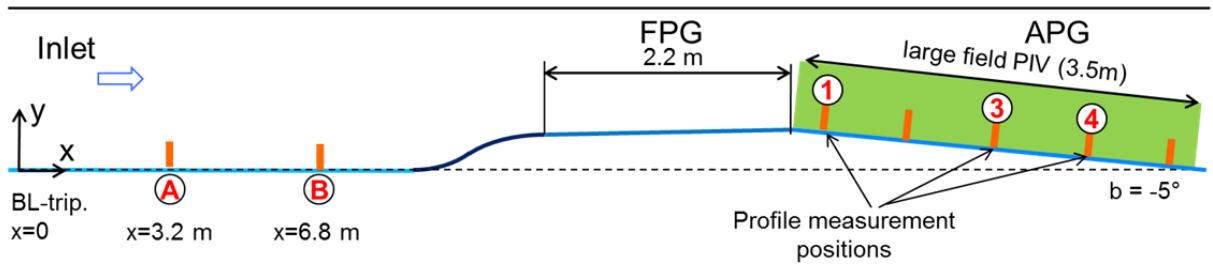

**Figure 10: Schematic of wind tunnel test section with APG-Model and profile measurement positions**

The 5 mm wide measurement area was illuminated by a pair of externally modulated continuous wave lasers (Kvant Laser, SK) with a combined output power of nearly 10W at a wavelength of 520 nm. Focusing lenses narrowed the 5 mm wide light sheet to a thickness of about 200-300 µm before passing through an anti-reflection-coated 3 mm thick glass window at the center of the acrylic glass panel (Figure 11, right).

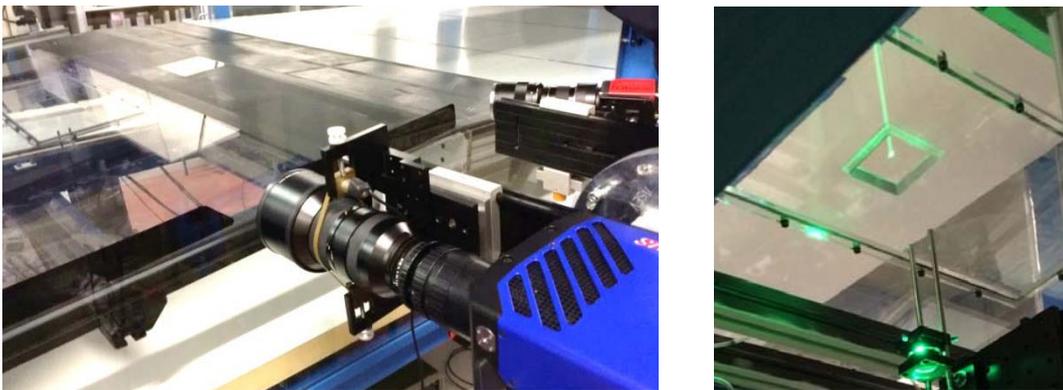

**Figure 11: Left: High-speed camera with 300mm lens. Right: laser light sheet introduced from below the wind-tunnel through an anti-reflection coated thin glass window.**

A telephoto lens (Zeiss 300mm/f2.8) with a 100 mm extension tube was chosen to image the near wall region with sufficiently high magnification (m=0.44) at a working distance of 1.1 m. Laser illuminated particle images were acquired using a high-speed CMOS camera (PCO, Dimax-S4) with a pixel pitch of 11 µm (corresponding to 25 µm/pixel in object space). Synchronization between modulated laser and camera was provided through a separate pulse control unit (Arduino MEGA 2560). At position (A), additional measurements were performed using a scientific-CMOS camera (PCO, Edge5.5) whose smaller pixel pitch of 6.5 μm permitted a correspondingly higher spatial resolution. In both cases the cameras were mounted with a 90° sideways rotation in order to achieve the highest possible sensor readout rates.

To account for the increase in magnification and reduction of the light sheet thickness, the seeding rate was increased by a factor of roughly 3 to 4 on the smoke generator compared to the large field PIV experiments.

At each measurement position, a minimum of two sequences consisting of more than 50,000 2D2C velocity field samples were acquired for each of the two free-stream flow conditions ($U_\infty$ = 5 m/s and 9 m/s). The image sequences were processed using a conventional 2-C PIV processing algorithm (PIVview 3.6, PIVTEC GmbH). Using a coarse-to-fine pyramid approach (Raffel, et al., 2007) with intermediate validation (normalized median filter and smoothing), the final interrogation window size is 64x8 or 32x8 pixels depending on the particle concentration, corresponding to 1.6x0.2 mm$^2$ or 0.8x0.2 mm$^2$ with an overlap of 75%. The large aspect ratio was chosen to achieve optimal wall-normal spatial resolution in the velocity profiles. Validation rates exceed 99.9% along the midline of the PIV recordings.

### 3.~~1~~3.4  Flow over the FPG region (FPG PIV).

The flow above the 1.5° flat plate was characterised through 2D2C PIV at several stations and for the two free-stream velocities ($U_\infty$ = 5 and 9 m/s). The aim was to characterise the accelerating flow upstream of the adverse pressure gradient (APG). The light sheet was introduced in the same way as for the large field 2D2C PIV in the APG region and the same optics was used. The mirror inside the wind tunnel in the downstream part of model was just raised by 50 cm. The light sheet was 1mm in thickness on the 2.14 m of the 1.5° plate. It was then adjusted tangent to the wall and normal to it in the mid-plane of the ramp. The experiment was conducted in two set-ups. The first one with two sCMOS cameras close to the end of the plate with an overlap of 10 mm between both fields of view, forming a global field of view of 46.5 cm long and 19.8 cm in height above the wall. This field of view was positioned to get a 7 mm overlap with the large field APG-TBL field of view. The second set-up used also two sCMOS cameras but without overlap between them. The field of view of each camera was 23.6 cm along the wall and 19.5 cm in the wall-normal direction. The first camera field of view starts at the beginning of the 1.5° plate and the second one is located close to the middle of this plate. For both cases, the cameras were equipped with Nikon 105 mm lenses at f# = 5.6. The time interval between frames was tuned to obtain 12 pixels displacements in the free-stream region to ensure sufficient velocity dynamic range for the turbulence intensity measurements. For the first set-up, 30 000 PIV fields were recorded for the two velocities and 10 000 for the second one. The data were processed in the same way as in the previous subsection by the modified version of the Matpiv toolbox by LML. The final interrogation window size corresponds to 2.25 by 2.25 mm in the physical space. The grid spacing was selected as 0.94 mm by 0.94 mm leading to an overlap of 58%.

| Experiment | Acronym | FOV (m x m) | Dt (px) | IW (px x px) | IW (mm x mm) | Vect field size | Number of samples | Frequency (Hz) |
|---|---|---|---|---|---|---|---|---|
| **Large field streamwise 2D2C PIV** | LFStW PIV | 3.466 x 0.255 | 14 | 24 x 24 | 2.56 x 2.56 | 3250 x 238 | 30 000 | 4 |
| **Spanwise SPIV** | SpW SPIV | 0.120 x 0.324 | 17 | 24 x 24 | 1.33 x 1.33 | 361 x 975 | 30 000 | 5 |
| **Flow over the FPG region part 1** | FPG PIV | 0.465 x 0.198 | 12 | 24 x 24 | 2.25 x 2.25 | 496 x 210 | 30 000 | 4 |
| **Flow over the FPG region part 2** | FPG PIV | 0.236 x 0.195 | 12 | 24 x 24 | 2.25 x 2.25 | 253 x 208 | 10 000 | 4 |
| **Time resolved High Magnification PIV PCO Dimax** | TRHM PIV | 0.0045 x 0.007 0.0045 x 0.025 | 25 | 32 x 8 | 0.8 x 0.2 | 5 x 144 5 x 504 | 63 464 251 583 | 667, 1000, 2000, 4000, 6667 |
| **Time resolved High Magnification PIV PCO edge** | TRHM PIV | 0.003 x 0.019 0.003 x 0.038 | 25 | 32 x 8 | 0.8 x 0.2 | 5 x 640 5 x 1280 | 10 000 50 000 | 100 200 |
| **Upstream BC** | UBC SPIV | 0.180 x 0.299 | 11 | 16 x 24 | 2.4 x 2.4 | 180 x 299 | 10 000 | 5 |
| **Corner Flow** | CF SPIV | 0.235 x 0.435 | 11 | 16 x 24 | 3.5 x 3.5 | 156 x 290 | 4 000 | 5 |

**Table 2: Main parameters of the PIV recordings from the different experiments.**

## 4. Pressure boundary conditions

Before the EuHIT experiment, the pressure distribution on the model was measured for two free-stream velocities at the wind tunnel entrance ($U_\infty$ = 5 and 9 m/s). This was done with the help of scannivalves and a Furness FCO 14 manometer having a range of 0 to 10 mm H$_2$O, and an uncertainty of ±0.5% of the reading. The pressure tap number 17 (located just before the 3.5 m plate) was chosen as reference as it corresponds to the smallest pressure on the model. The pressure coefficient $Cp = \dfrac{P - P_{17}}{\frac{1}{2}\rho U_\infty^2}$ was computed with $P_{17}$ being the reference pressure, 'ρ' the air density and '$U_\infty$' the free stream velocity upstream of the ramp at 10 cm downstream of test section entrance. The uncertainty on $C_p$, deduced from the Cp distribution is 1.2% of the local value, and 6.5 % on the pressure gradient (see (Cuvier, et al., 2014)).

Figure 12 gives the Cp distribution along the model for two free stream velocities (5 and 9 m/s) while Figure 13 gives the corresponding pressure gradient distributions. As can be seen in Figure 12, the flow accelerates in the converging part 0 ≤ s ≤ 1360 mm of the ramp causing a decrease in the pressure coefficient until the suction peak at s=1146 mm. This suction peak then induces a locally strong adverse pressure gradient (see Figure 13). Behind this suction peak, a region of pressure recovery develops until the flow begins to accelerate again due to the favourable pressure gradient caused by the 2.14 m long flat plate inclined at 1.5°. A second suction peak is observed at the articulation of the ramp (s=3500 mm) close to the reference pressure tap 17. Then the 3.5 m long ramp causes a region of relatively constant adverse pressure gradient with a pressure gradient coefficient β about 2.

To be sure that the 45° mirror inside the tunnel and the slot in the wall had limited effects on the flow upstream, the streamwise pressure distribution along the ramp was also acquired with the mirror and the slot. No difference was observed at least not until the last pressure tap on the model (which is 460 mm before the end of the ramp).

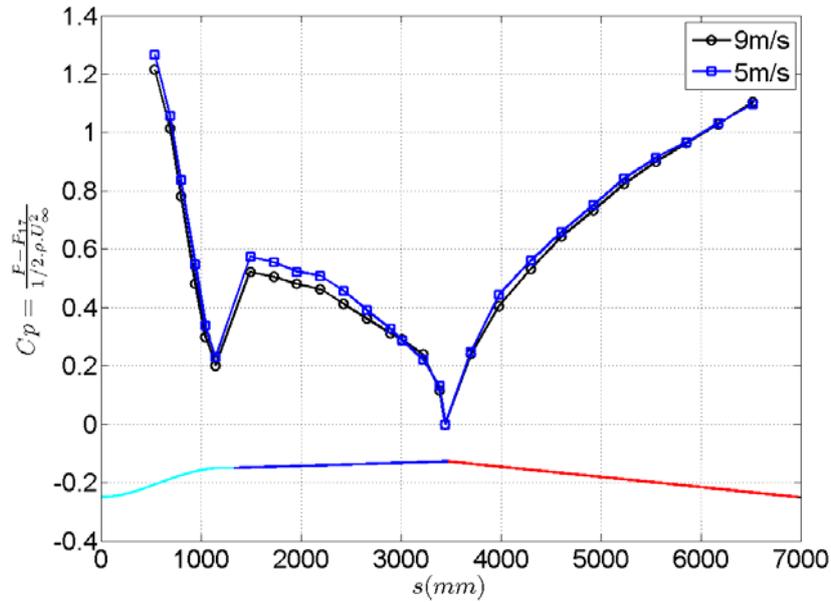

**Figure 12: Streamwise Pressure Coefficient distribution along the ramp for $U_\infty$ = 5 and 9 m/s**

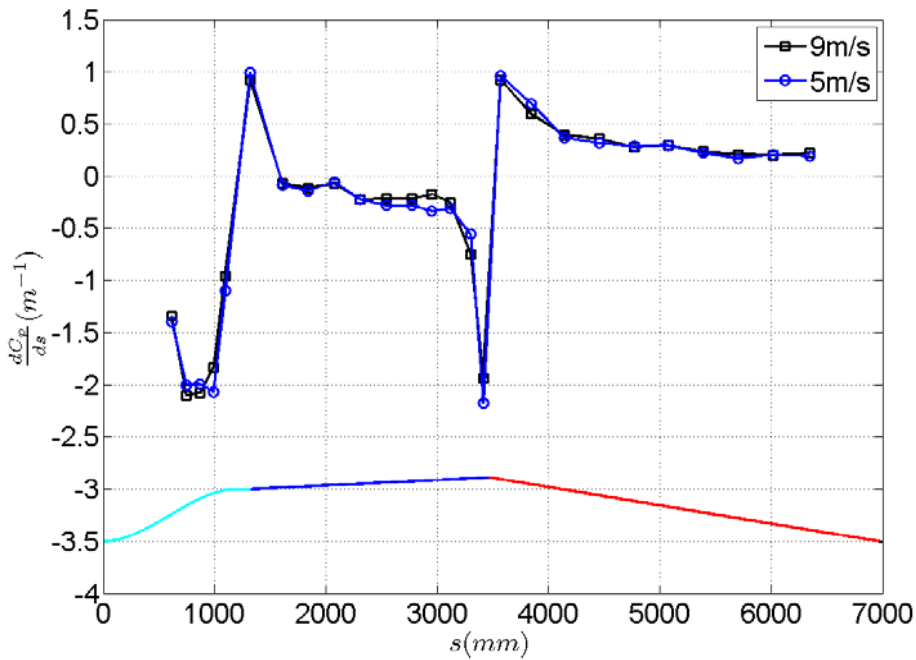

**Figure 13: Streamwise Pressure gradient distribution along the ramp for $U_\infty$ = 5 and 9 m/s**

Spanwise pressure measurements were also undertaken at four locations to check the spanwise homogeneity and the two dimensionality of the flow. The first two stations were located on the 1.5° plate at s=1727 mm and s= 3010 mm while the last two stations were located on the -5° plate at s=4301 mm and s=6176 mm. As seen in Figure 14 and Figure 15, the pressure distribution is almost constant at the 4 stations for both external velocities except slight variations close to z = ±800 mm. This confirms the two dimensionality of the flow for -600 ≤ z ≤ 600 mm.

To provide complete boundary conditions, the pressure distribution on the wind tunnel top wall was measured using a plate equipped with 3 pressure taps. The distribution was acquired by moving this plate along the upper wall. The effect of the ramp becomes visible around X= 8 m (see Figure 16). Mean velocity and Reynolds stresses profiles at X = 6.8 m (2.6 m upstream of the ramp leading edge), which can serve as inlet boundary conditions for CFD, are then not affected by the presence of the model. The smoothness of the Cp curve shows the very good repeatability of the flow. The pressure distribution starts with a slight favourable pressure gradient (see Figure 17) due to the boundary layer development on each wall of the test section. It is -0.44 Pa/m for $U_\infty$ = 9 m/s and -0.19 Pa/m for $U_\infty$ = 5 m/s close to the original values found by (Carlier & Stanislas, 2005). Then a trend similar to the pressure distribution on the model is observed. The presence of the contraction and the flat plate at 1.5° creates a flow acceleration which induces a strong favourable pressure gradient until X = 12.54 m after which the flow switches to an adverse pressure gradient caused by the 3.5 m ramp.

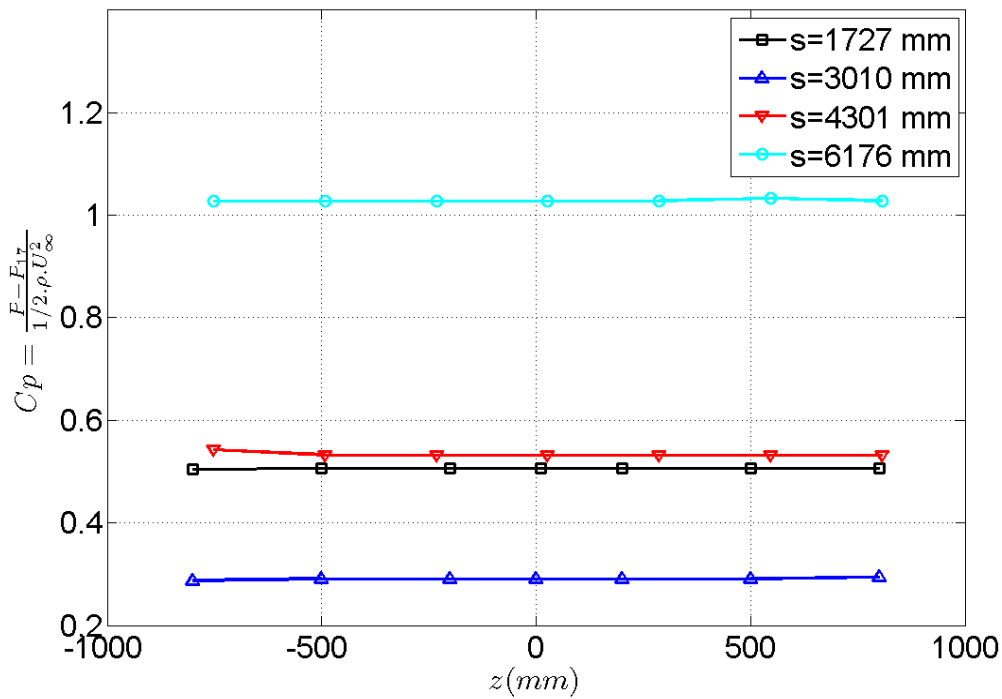

**Figure 14: Spanwise Pressure coefficient distribution along the ramp for $U_\infty$ = 9 m/s**

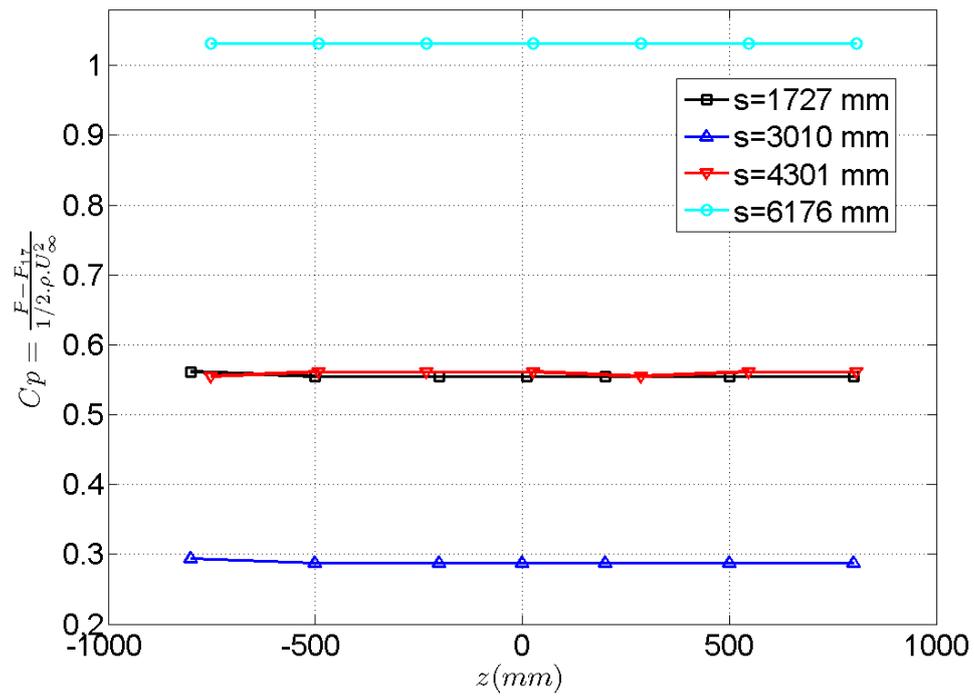

**Figure 15: Spanwise Pressure Coefficient distribution along the ramp for U$_\infty$ = 5 m/s**

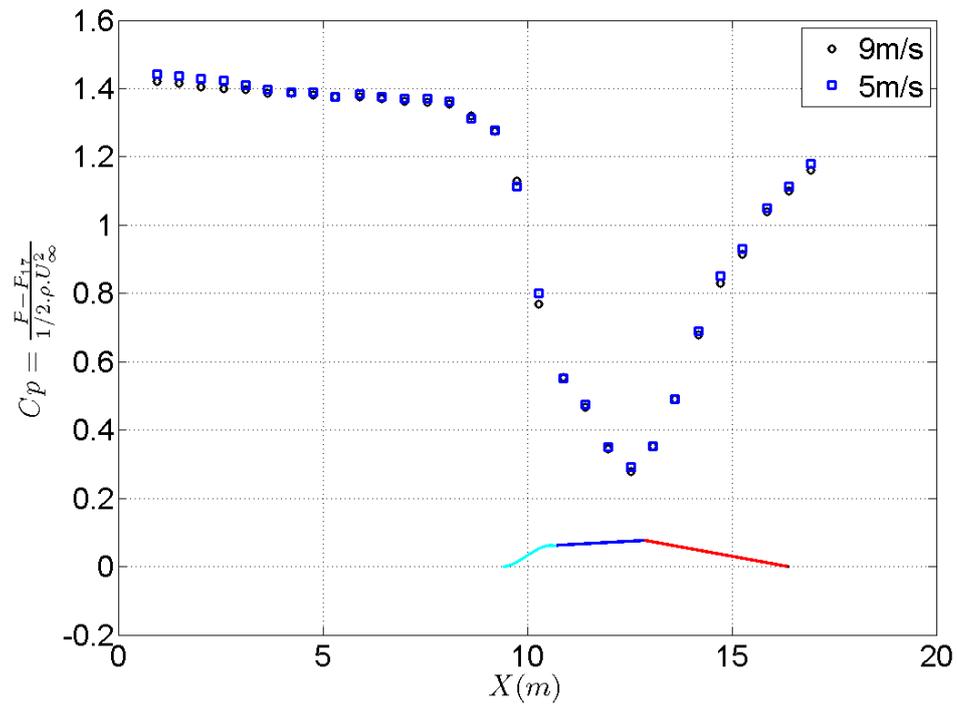

**Figure 16: Streamwise Pressure Coefficient distribution along the roof for U$_\infty$ = 5 and 9 m/s**

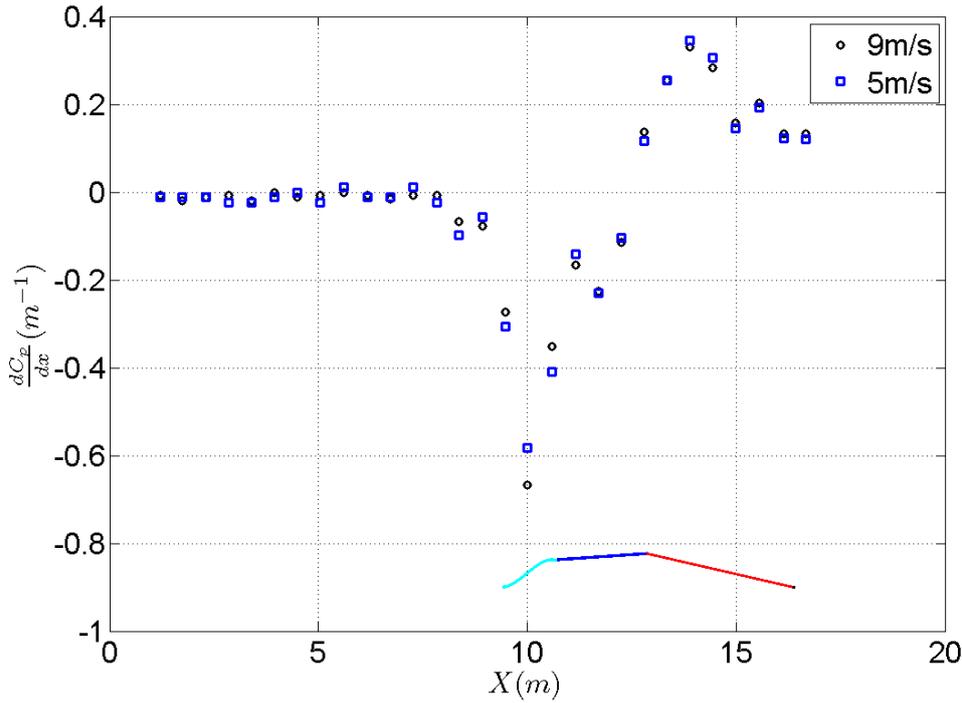

Figure 17: Streamwise Pressure gradient distribution along the roof for $U_\infty$ = 5 and 9 m/s

## 5. Characterisation of the flow in the FPG part of the ramp.

In order to provide a detailed characterization of the flow over the whole model PIV measurements were also performed with the set-up described in section 3.4 of the flow over the converging flat part of the model. The local coordinate attached to the ramp is used again. The s direction is the direction parallel to the 1.5° ramp, y is the wall-normal direction and z the transverse one, with z = 0 corresponding to the wind tunnel centreline. Figure 18 shows the evolution of the mean streamwise velocity profiles for both velocities along the 1.5° plate normalised with the reference velocity $U_\infty$ and $\delta_0$ at s= 1.362 m (which is 101 mm at 5 m/s and 95 mm at 9 m/s). The first profile at s = 1.362 m (just after the beginning of the 1.5° plate) exhibits a small peak close to $y/\delta_0$ = 0.15. The external velocity is about 1.4 $U_\infty$ due to the preceding contraction. At the second station the peak is still visible but strongly attenuated and further away from the wall ($y/\delta_0$ = 0.4). After this station, the peak disappears. From the beginning of the plate to its end, the mean velocity is increasing continuously due to the favourable pressure gradient (FPG) encountered, with an external velocity reaching 1.55 $U_\infty$ at the last station. At this last station, which is very close to the ramp articulation, the acceleration is highly marked due to the strong FPG generated by the sudden change of flow direction. One should mention that, for each profile, the first PIV measurement point above the wall is biased due to wall proximity so it was suppressed. Table 3 and Table 4 give the boundary layer characteristics on the 1.5° plate at the same stations as in Figure 18 for $U_\infty$ = 5 m/s and 9 m/s respectively. The evolution of the shape factor indicates a strong similarity of behaviour of the boundary layer at the two Reynolds numbers in this region of the flow.

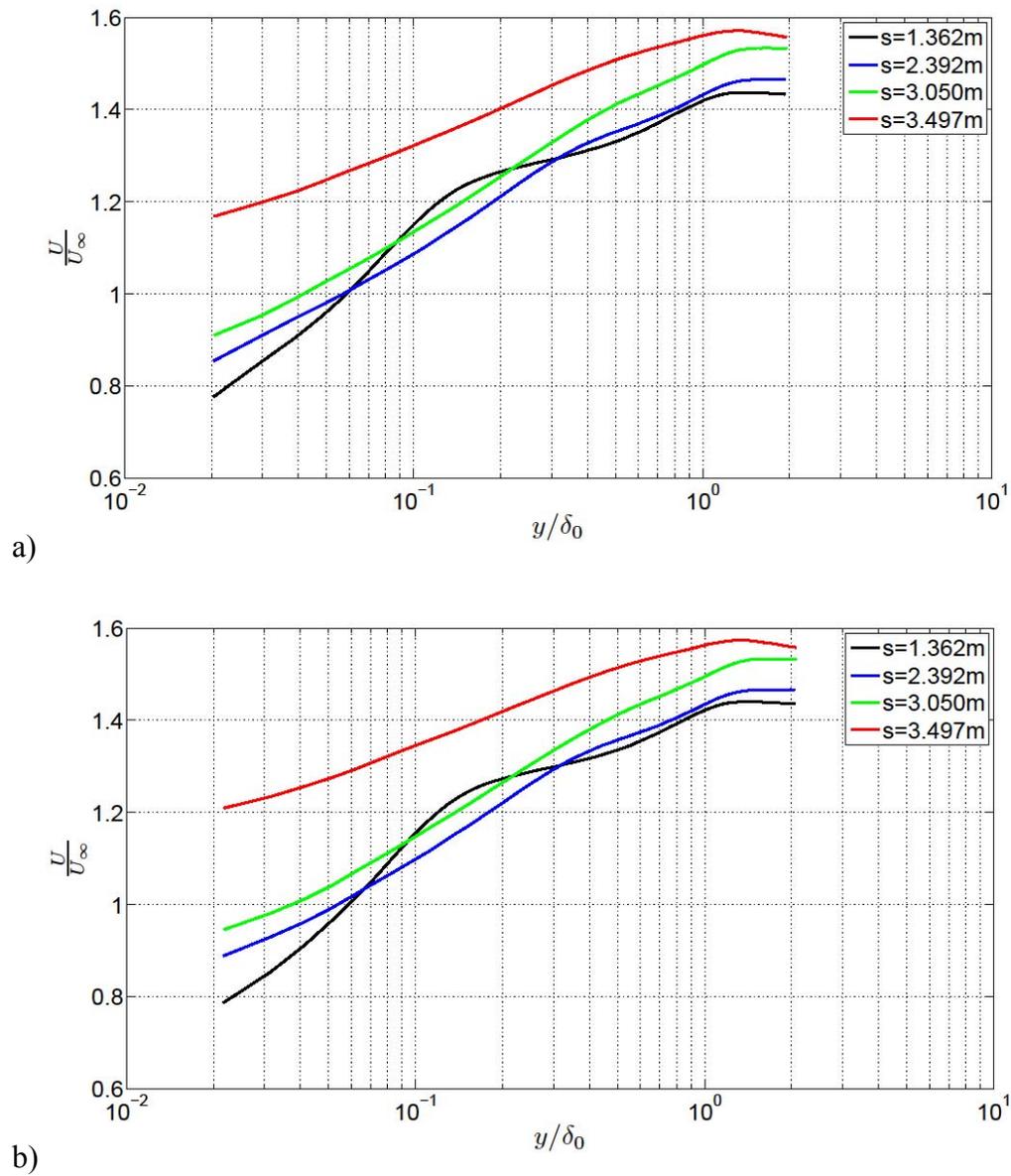

a)

b)

**Figure 18:** Evolution of the mean streamwise velocity profiles along the 1.5° plate, a) $U_\infty$=5m/s, b) $U_\infty$ = 9m/s. δ is taken at s= 1.362 m.

| Station s (m) | $U_e$ (m/s) | δ (mm) | δ* (mm) | θ (mm) | H | $Re_\theta$ | β |
|---|---|---|---|---|---|---|---|
| 1.362 | 7.17 | 101 | 9.8 | 7.6 | 1.30 | 3630 | 1.28 |
| 2.392 | 7.33 | 118 | 12.1 | 9.6 | 1.26 | 4680 | -0.42 |
| 3.050 | 7.68 | 117 | 12.2 | 9.6 | 1.27 | 4920 | -0.50 |
| 3.497 | 7.80 | 78 | 05.5 | 4.3 | 1.28 | 2230 | -0.32 |

**Table 3 :** Boundary layer characteristics at different stations on the 1.5° plate at $U_\infty$ = 5m/s.

| Station s (m) | $U_e$ (m/s) | $\delta$ (mm) | $\delta^*$ (mm) | $\theta$ (mm) | H | $Re_\theta$ | $\beta$ |
|---|---|---|---|---|---|---|---|
| 1.362 | 12.93 | 95 | 9.0 | 7.0 | 1.29 | 6000 | 1.41 |
| 2.392 | 13.20 | 110 | 11.0 | 8.7 | 1.26 | 7630 | -0.37 |
| 3.050 | 13.79 | 113 | 11.3 | 8.9 | 1.27 | 8170 | -0.33 |
| 3.497 | 14.03 | 71 | 4.5 | 3.5 | 1.29 | 3280 | -0.24 |

**Table 4: Boundary layer characteristics at different stations on the 1.5° plate at $U_\infty$ = 9 m/s.**

Figure 19 shows the evolution of the streamwise turbulence intensity profiles for both velocities along the 1.5° plate. For both velocities the standard near wall peak is not accessible as the first measurement point is too far away from the wall. The first profile at s = 1.362 m shows a small plateau close to $y/\delta_0$ = 0.03-0.05. This plateau was also observed by (Cuvier, et al., 2014) for a different configuration of the model. It is probably due to the strong and short APG encountered at the end of the converging part of the model (see Figure 13). This plateau remains as a kink in the profile, away from the wall, at the three following stations ($y/\delta_0 \sim$ 0.1, 0.15 and 0.2 successively). In the outer part, a second plateau is clearly visible in the first profile, close to $y/\delta_0$ = 0.5, which progressively transforms into a vanishing kink at the following stations. This second plateau is probably linked to the external turbulence of the incoming boundary layer which is seriously attenuated by the strong FPG encountered in the converging part of the ramp. At s = 3.497 m, both kinks are nearly smoothed out.

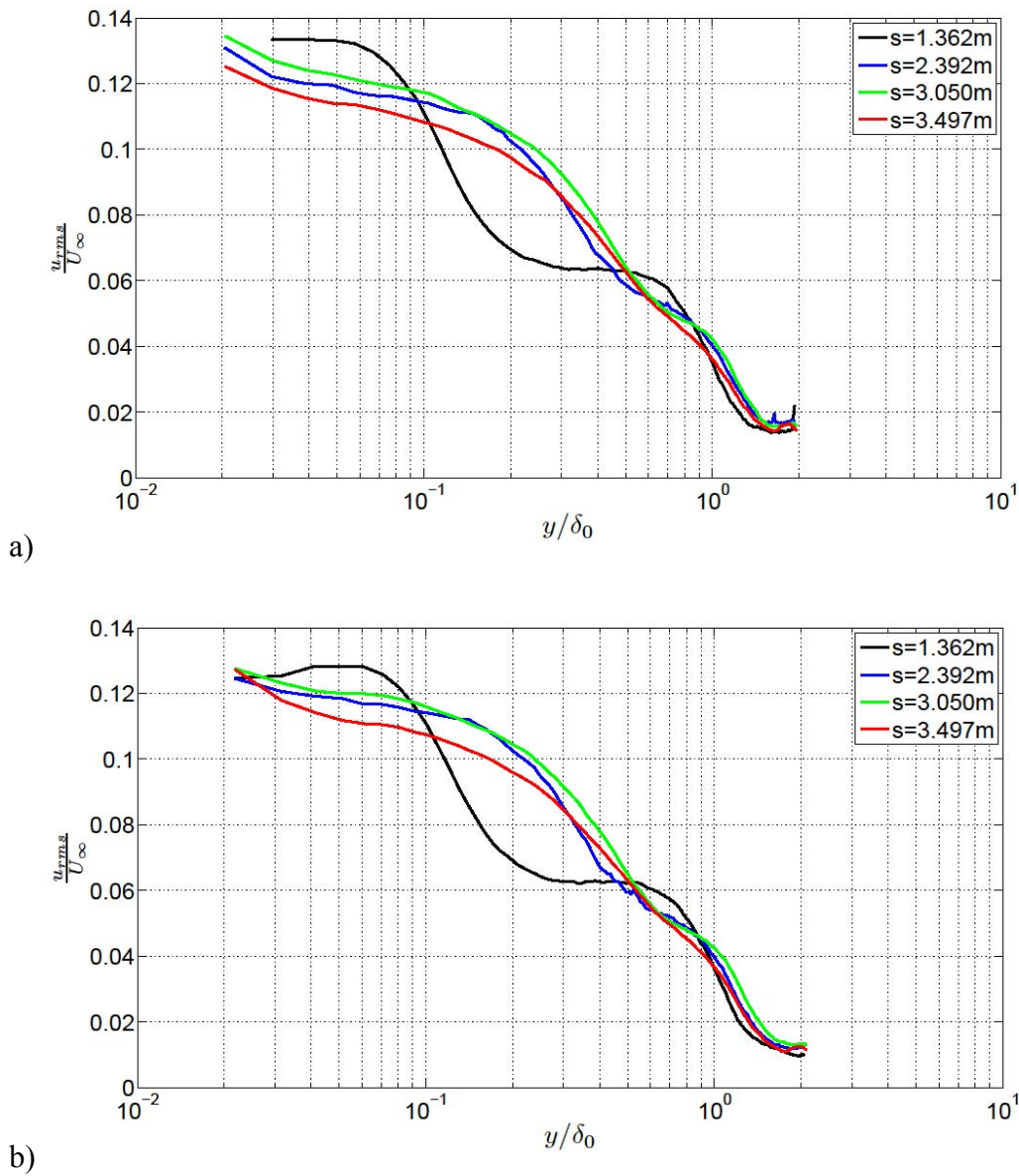

a)

b)

**Figure 19: Evolution of the streamwise turbulence intensity profiles along the 1.5°, a) $U_\infty$=5m/s, b) $U_\infty$ = 9m/s. $\delta_o$ is taken at s= 1.362 m.**

Figure 20 shows the evolution of the wall-normal turbulence intensity profiles for both velocities. At the first station, the shape is similar to the streamwise component but the plateaux are replaced by well defined peaks. At the following stations, the outer peak spreads out and nearly disappears while the one closer to the wall stays fairly visible.

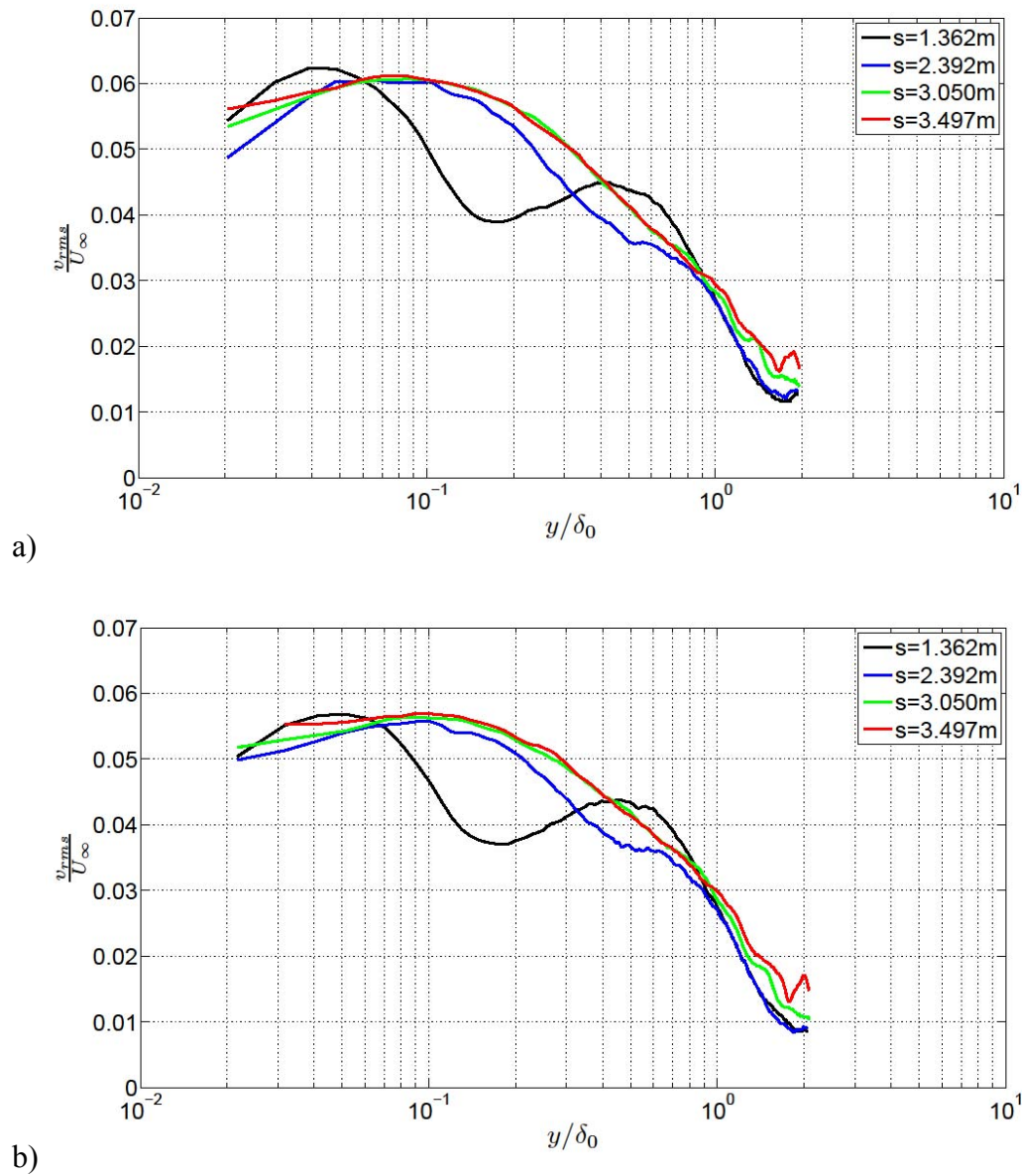

a)

b)

**Figure 20: Evolution of the wall-normal turbulence intensity profiles along the 1.5° plate, a) $U_\infty$=5m/s, b) $U_\infty$ = 9m/s. $\delta_o$ is taken at s= 1.362 m.**

Figure 21 shows the evolution of the Reynolds shear stress profiles for both velocities. The behaviour is obviously strongly influenced by the v component of the fluctuations (see Figure 20).

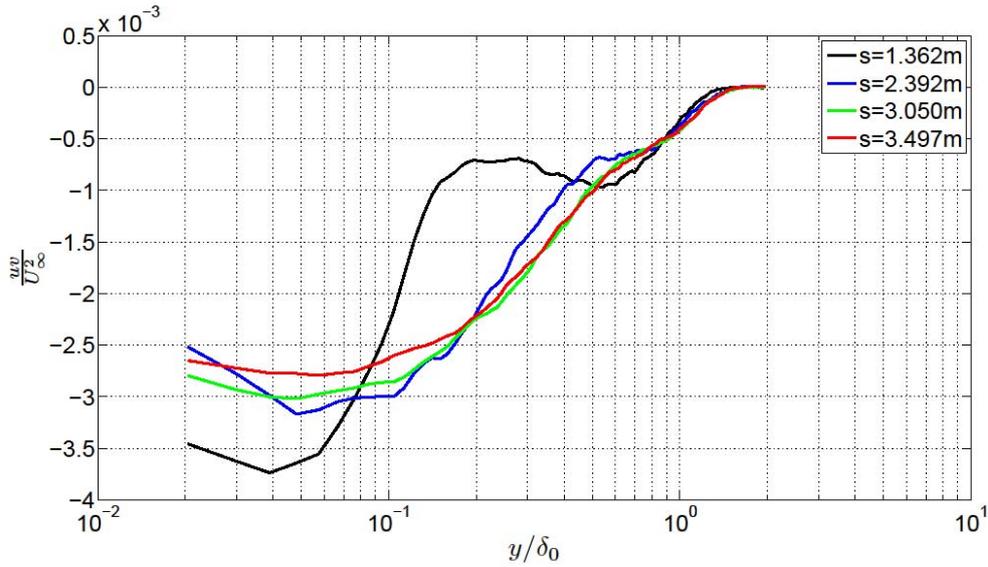

a)

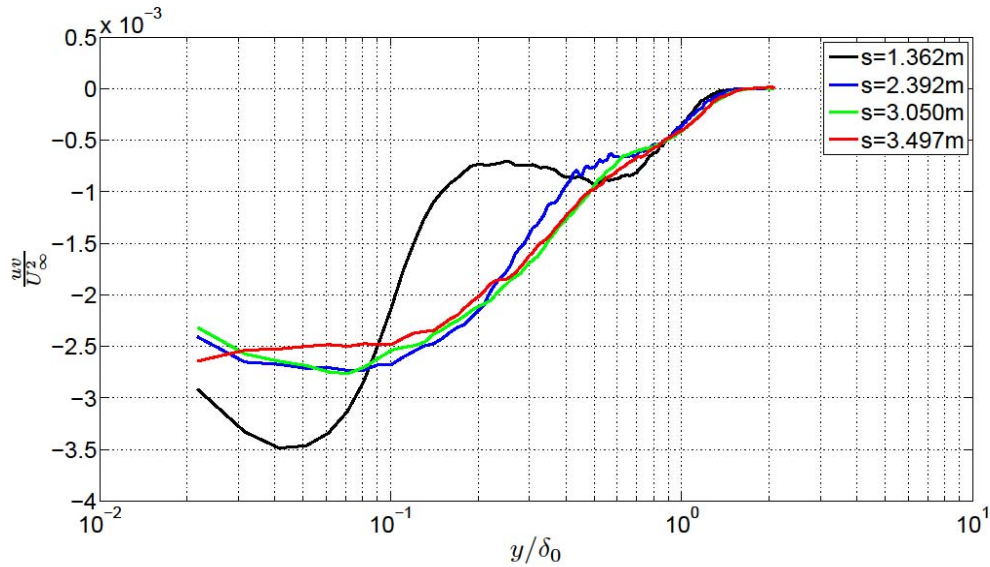

b)

**Figure 21: Evolution of the Reynolds shear stress profiles along the 1.5° plate, a) $U_\infty$=5m/s, b) $U_\infty$ = 9m/s. $\delta_o$ is taken at s= 1.362 m.**

## 6. Characterization of the APG flow

This section gives the main results obtained with the very large fields of view acquired with 16 sCMOS cameras as described in section 3.1. The aim is to provide detailed statistics of the flow in the APG region for numerical simulations comparisons. Figure 22 shows the evolution the mean streamwise velocity normalised with the reference free-stream velocity $U_\infty$ along the -5° plate for both velocities studied. Due to test section enlargement, the flow decelerates continuously and the thickness of the boundary layer increases. The ratio between the local free-stream velocity at the beginning and at the end of the field is 1.45 on 3.5 m. As expected, the moderate APG imposed on this part of the model gives no flow separation.

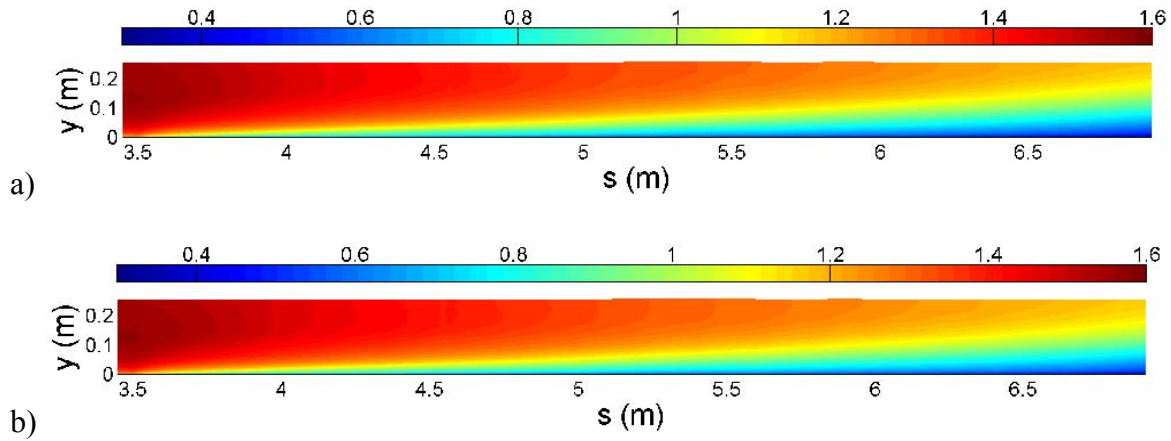

**Figure 22: Mean streamwise velocity U/ $U_\infty$ fields in the APG region for the two free stream velocities, a) $U_\infty$ = 5 m/s, b) $U_\infty$ = 9 m/s.**

Figure 23 shows the evolution of the streamwise turbulence intensity field along the APG plate for both velocities studied and normalized with $U_\infty$. A good continuity is observed between the 16 fields of view assembled in this set-up. As observed in all APG flows with and without separation ( (Simpson, 1989), (Webster, et al., 1996), (Wu & Squires, 1998), (Cuvier, et al., 2014)), a region of high values reaching about 13% of $U_\infty$ develops above the wall which both spreads and moves away from it when developing downstream.

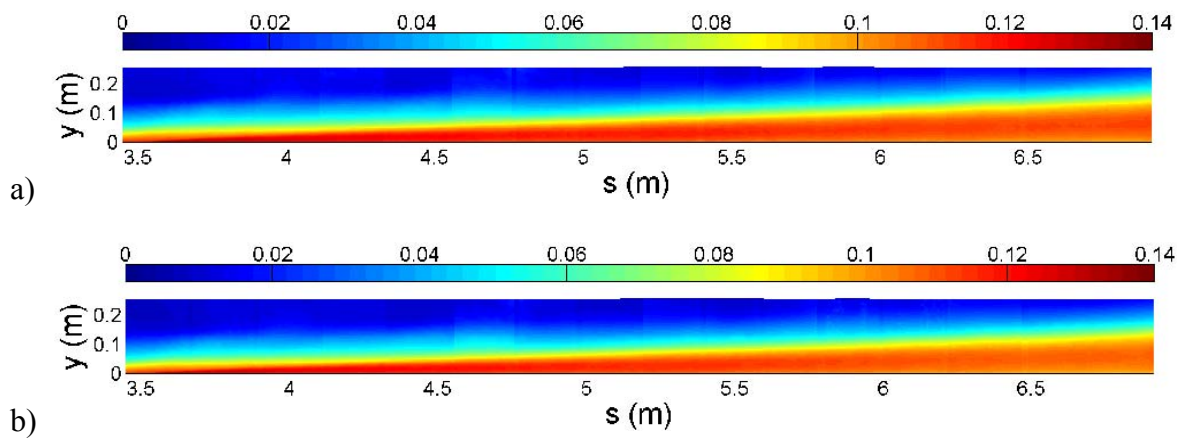

**Figure 23: Streamwise turbulence intensity $u_{rms}$/ $U_\infty$ fields in the APG region for the two free stream velocities, a) $U_\infty$ = 5 m/s, b) $U_\infty$ = 9 m/s.**

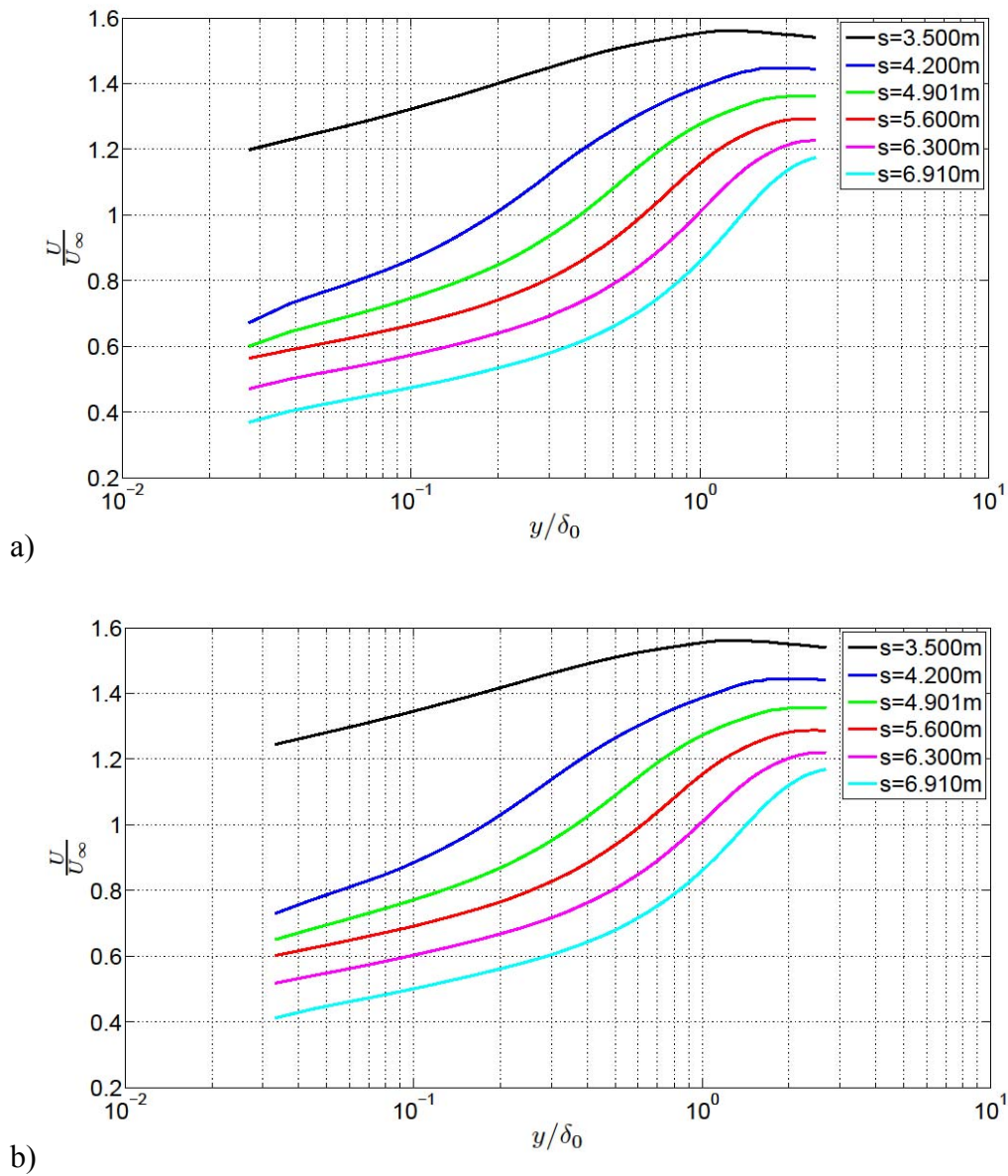

**Figure 24: Evolution of the mean streamwise velocity profiles along the -5° plate, a) $U_\infty$ = 5m/s, b) $U_\infty$ = 9m/s. $\delta_o$ is taken at s= 3.5 m.**

To get a more quantitative view on the evolution of the mean streamwise velocity, Figure 24 shows profiles for both free stream velocities along the APG plate normalised by $U_\infty$ and $\delta_0$ at s = 1.362 m ($\delta_0$ is 101 mm at 5 m/s and 95 mm at 9 m/s). In the first profile at s = 3.500 m, just at the start of the ramp, the velocity is globally higher than the free stream. It is very comparable to the last profile of the preceding FPG plate. Then progressively the APG reduces globally the velocity with a stronger effect near the wall. The similarity of the profiles between the two Reynolds numbers is interesting to note. Table 5 and Table 6 give the main boundary layer parameters at the six stations presented in Figure 24. Although the profiles are quite similar between the last station of the FPG and the first station of the APG, a significant difference appears in the value of the shape factor, indicating that the strong variations of the pressure gradient observed in Figure 13 close to the articulation of the APG ramp affects significantly the near wall region. Downstream, the reduction of the external velocity by a factor 1.3 along the 3.5 m of the plate is accompanied by a significant increase of the boundary layer thickness (a factor nearly 4) and of the integral parameters. Notably, the momentum thickness increases by a factor larger than 10, indicating that the boundary layer is moving toward separation (but not reaching it). This is confirmed by the values of the shape factor at the last station which are not far from the classical separation criteria. Subtle differences are observed between the two Reynolds numbers on the shape factor evolution which should be a modelling challenge.

| Station s (m) | $U_e$ (m/s) | $\delta$ (mm) | $\delta^*$ (mm) | $\theta$ (mm) | H | $Re_\theta$ | $\beta$ |
|---|---|---|---|---|---|---|---|
| 3.500 | 7.70 | 66.0 | 3.9 | 2.9 | 1.34 | 1470 | -0.26 |
| 4.200 | 7.22 | 135.0 | 18.7 | 13.4 | 1.40 | 6430 | 1.42 |
| 4.901 | 6.81 | 161.0 | 26.1 | 18.0 | 1.45 | 8180 | 2.04 |
| 5.600 | 6.47 | 183.0 | 33.9 | 22.8 | 1.49 | 9840 | 2.01 |
| 6.300 | 6.13 | 205.0 | 44.2 | 28.4 | 1.56 | 11600 | 3.68 |
| 6.910 | 5.87 | 231.0 | 58.7 | 35.2 | 1.67 | 13790 | - |

Table 5: Boundary layer parameters in the APG region at $U_\infty$ = 5 m/s.

| Station s (m) | $U_e$ (m/s) | $\delta$ (mm) | $\delta^*$ (mm) | $\theta$ (mm) | H | $Re_\theta$ | $\beta$ |
|---|---|---|---|---|---|---|---|
| 3.500 | 13.86 | 57.0 | 2.8 | 1.9 | 1.47 | 1720 | -0.19 |
| 4.200 | 12.96 | 128.0 | 16.7 | 12.1 | 1.38 | 10490 | 1.53 |
| 4.901 | 12.20 | 154.0 | 23.3 | 16.5 | 1.41 | 13420 | 1.94 |
| 5.600 | 11.59 | 175.0 | 30.5 | 21.0 | 1.45 | 16240 | 2.27 |
| 6.300 | 10.97 | 196.0 | 39.6 | 26.3 | 1.51 | 19230 | 3.74 |
| 6.910 | 10.51 | 226.0 | 53.7 | 33.4 | 1.61 | 23430 | - |

Table 6: Boundary layer parameters in the APG region at $U_\infty$ = 9 m/s.

Figure 25 shows the evolution of the streamwise turbulence intensity profile in the APG region at the same stations as in Figure 24. The profiles at the first station is very similar to the one at the last station of the FPG (see Figure 19) with a slight reduction of level which is more marked near the wall at the highest Reynolds number. At the second station (s = 4.2m), a weak second peak, located close to $y/\delta_0$ = 0.15, is induced by the change of sign of the pressure gradient (the first near wall peak is not captured as it is too close to the wall even for the 5 m/s case). Downstream, this peak moves away from the wall and decreases slightly with increasing s. At the last station it has reached $y/\delta_0$ = 0.7 which corresponds to about 30% of the local boundary layer thickness $\delta$.

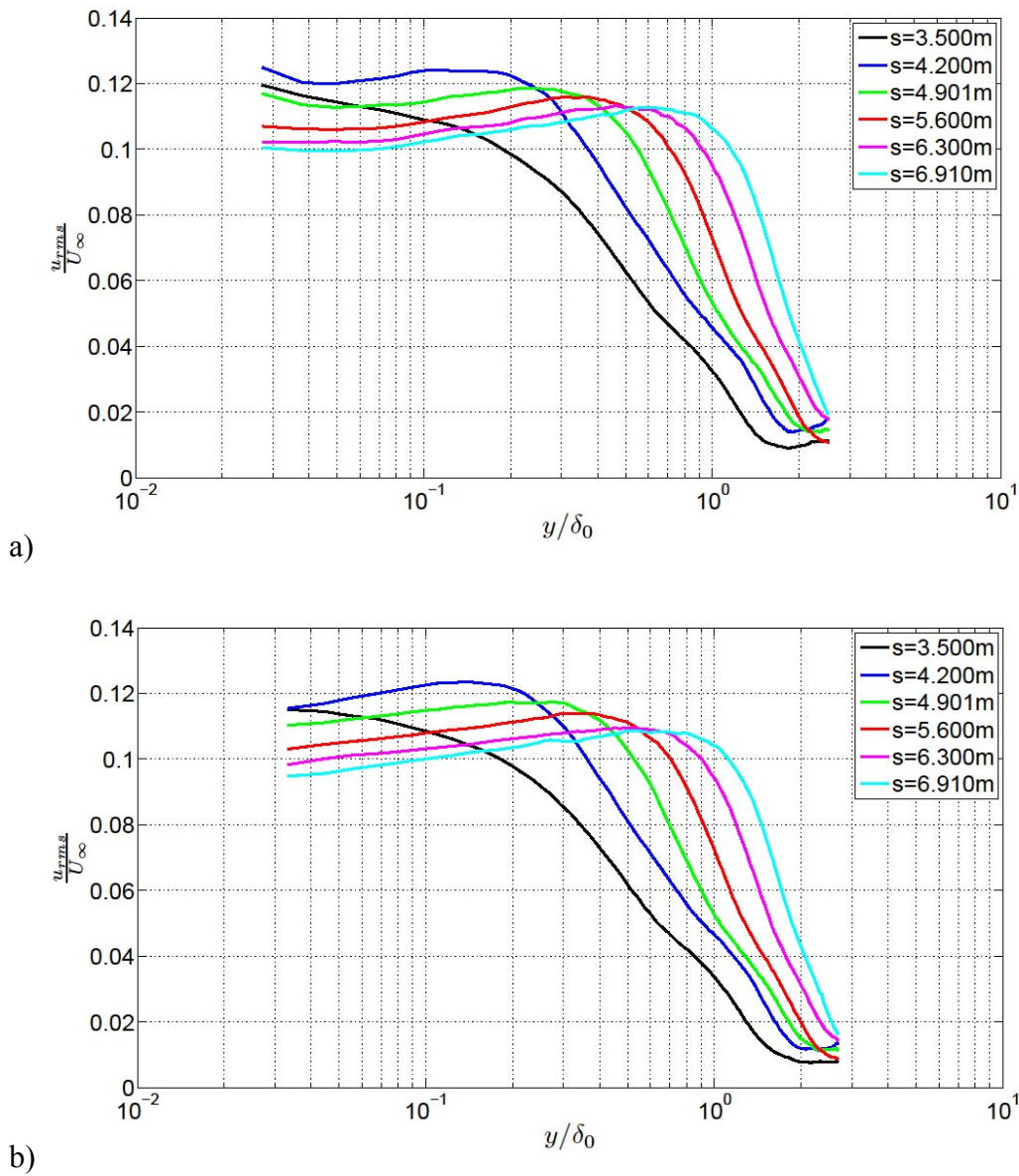

**Figure 25: Evolution of the streamwise turbulence intensity profiles along the -5° plate, a) $U_\infty$ = 5m/s, b) $U_\infty$ = 9m/s. $\delta_o$ is taken at s= 3.5 m.**

Figure 26 shows the evolution of the wall-normal turbulence intensity at the same stations. As for the streamwise component, the APG induces a second weak peak which move away from the wall with increasing s. Compared to the one observed on $u_{rms}$, the position with respect to the wall is very similar at the different stations but the height of this peak seems to increase with s especially for the highest velocity studied. This behavior was also observed by (Cuvier, et al., 2014) for an APG flow with separation.

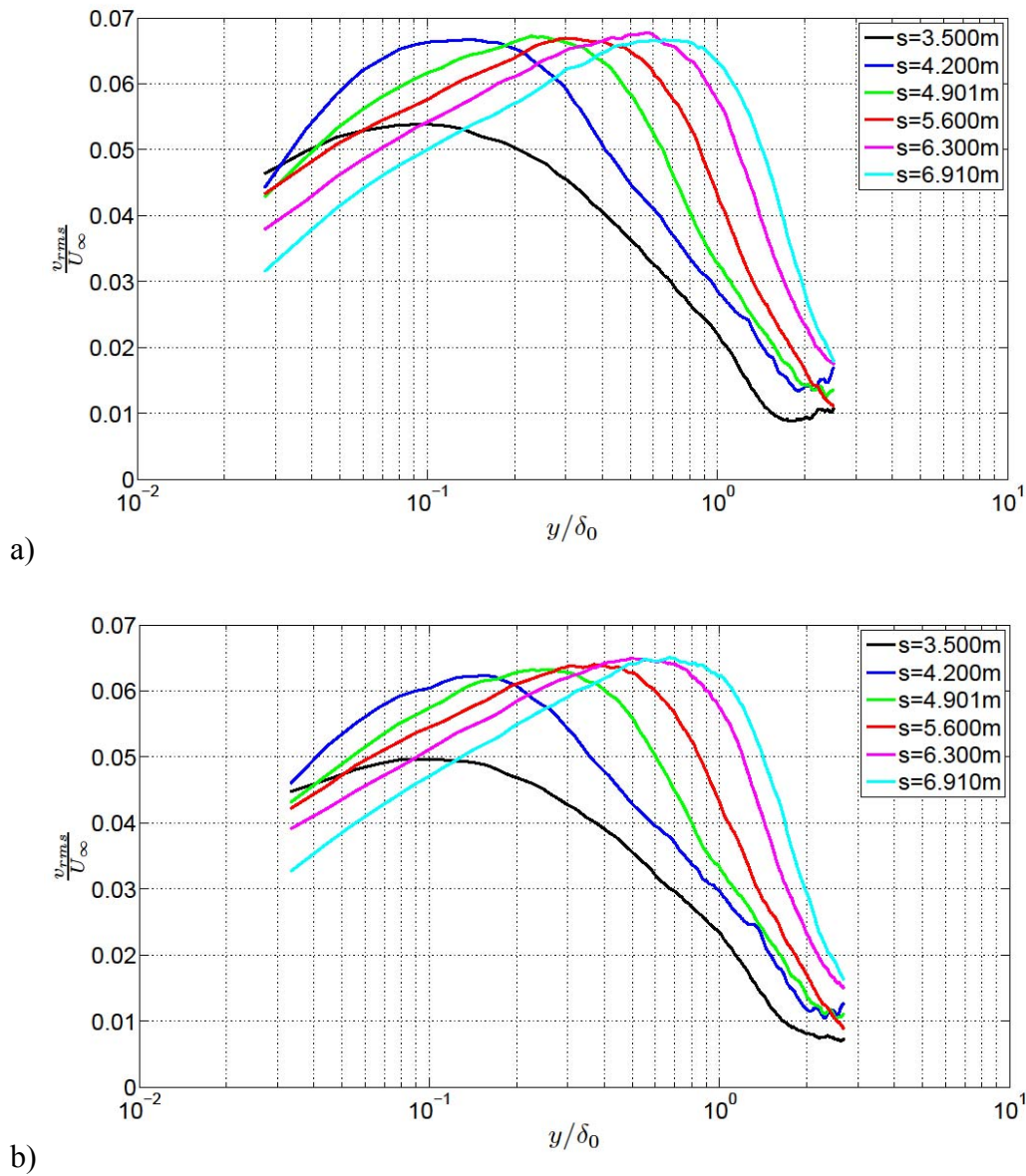

a)

b)

**Figure 26: Evolution of the wall-normal turbulence intensity profiles along the -5° plate, a) $U_\infty$= 5m/s, b) $U_\infty$ = 9m/s. $\delta_o$ is taken at s= 3.5 m.**

Figure 27 shows the Reynolds shear stress profiles at the same stations of the APG region. As for the streamwise and wall normal ones, this quantity presents a weak negative peak which moves away from the wall with s. Similar to the streamwise fluctuation, the amplitude of this peak decreases downstream.

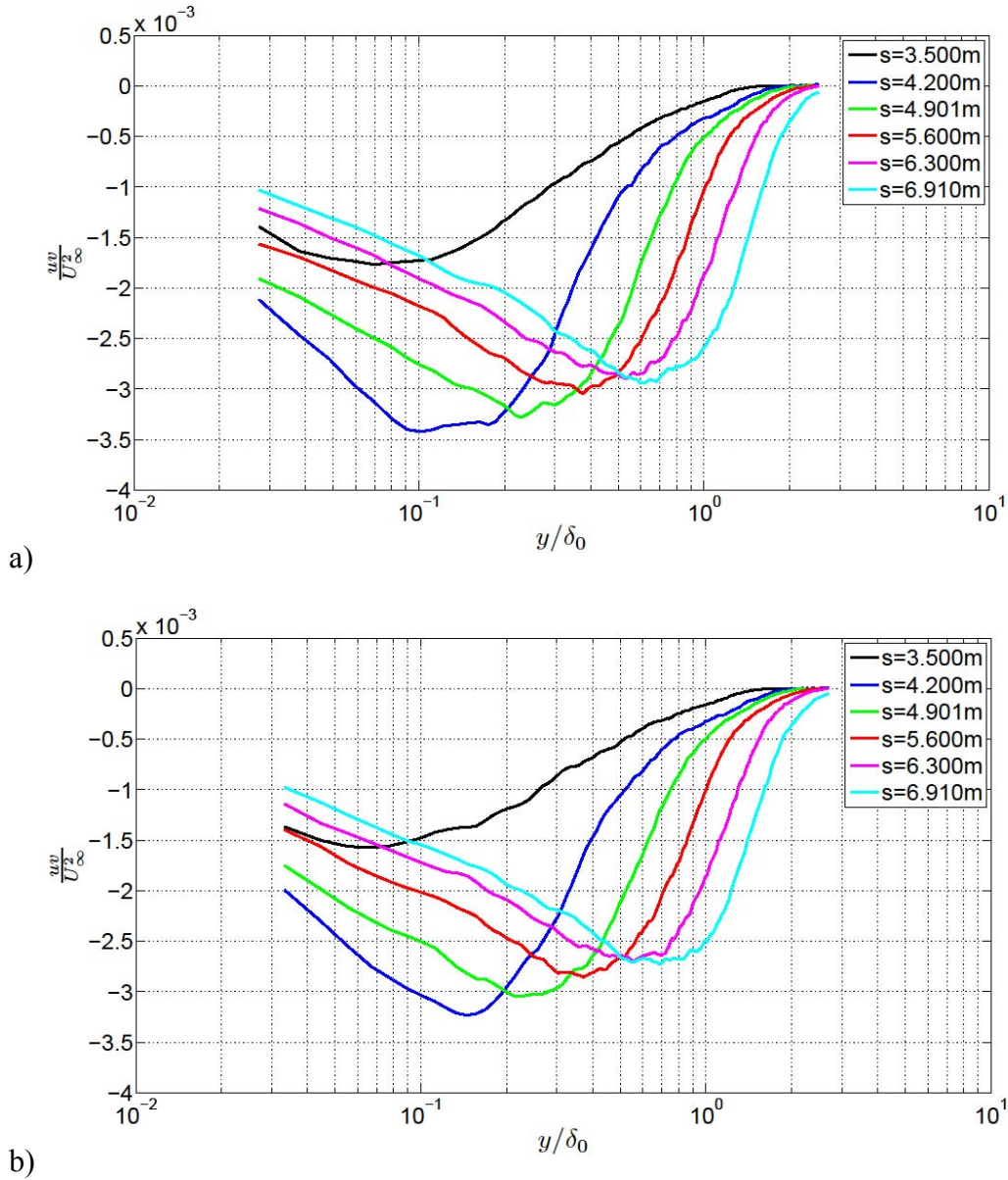

**Figure 27: Evolution of the Reynolds shear stress profiles along the -5° plate, a) $U_\infty$= 5m/s, b) $U_\infty$ = 9m/s. $\delta_o$ is taken at s= 3.5 m.**

TRHM PIV measurements were conducted at three stations in the APG region (s = 3.983, 5.233 and 5.858 m). Figure 28 to Figure 33 give the full mean streamwise velocity, streamwise and wall-normal turbulence intensity and Reynolds shear stress profiles at these three stations. The near wall part and the friction velocity are obtained from the TRHM PIV (section 3.3) and the outer part from the LFStW PIV (section 3.1). One should first note the good agreement between the two methods. It is then interesting to observe that a logarithmic region exist on all profiles but with a very limited extent. Also, the high magnification measurements allow to resolve the near wall peak of the streamwise turbulent fluctuations which is at its usual location (y+ ~ 15) and clearly marked at the three stations. The presence of this peak is an indication of the existence of near wall streaks all along the APG region. In this wall units representation, the inner peak is nearly constant, despite the pressure gradient, but the outer peak is growing to reach a level comparable to the near wall one at the last station. On the two other stress components, only the outer peak is clearly visible and it is also growing downstream. As several points are inside the viscous sublayer, the TRHM PIV measurements allow estimating the friction velocity which is given in Table 7 for both Reynolds numbers.

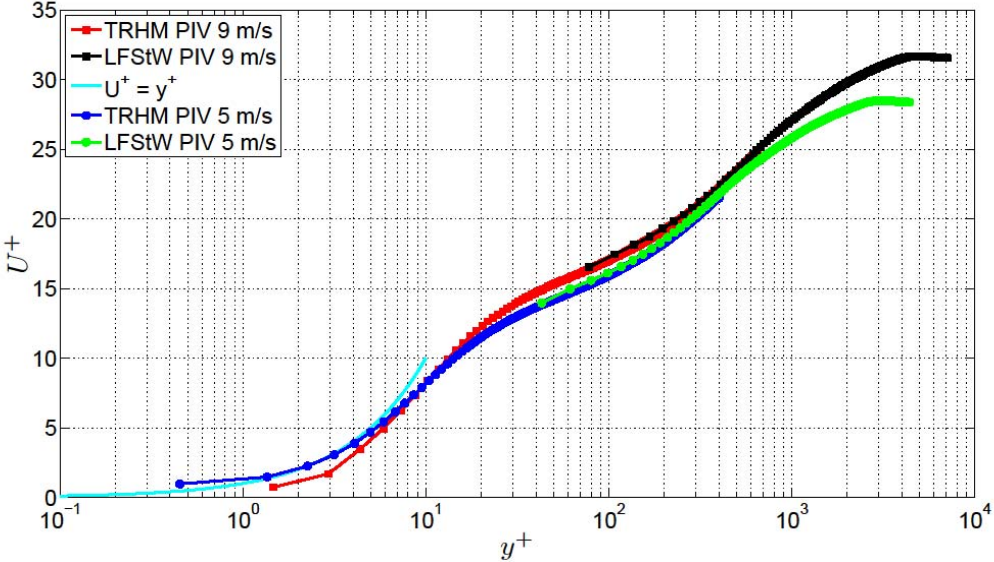

**Figure 28: Mean streamwise velocity profile at s = 3.983 m in the APG region for both Reynolds numbers.**

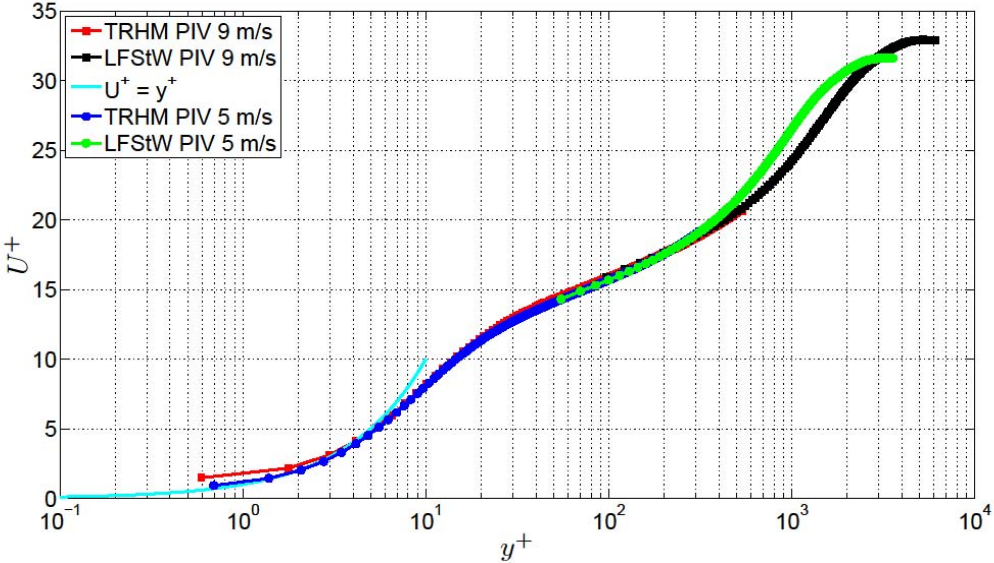

**Figure 29: Mean streamwise velocity profile at s = 5.233 m in the APG region for both Reynolds numbers.**

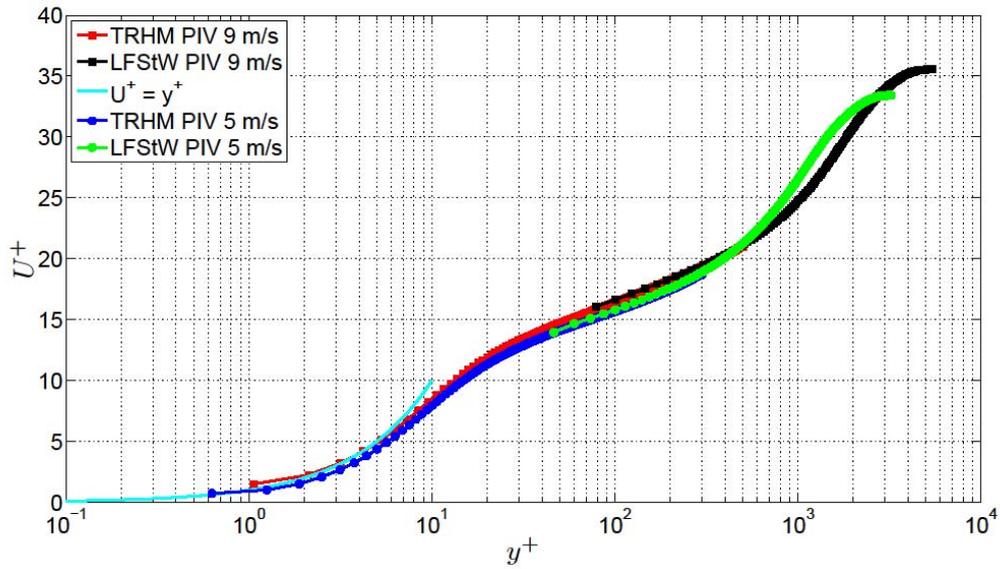

Figure 30: Mean streamwise velocity profile at s = 5.858 m in the APG region for both Reynolds numbers.

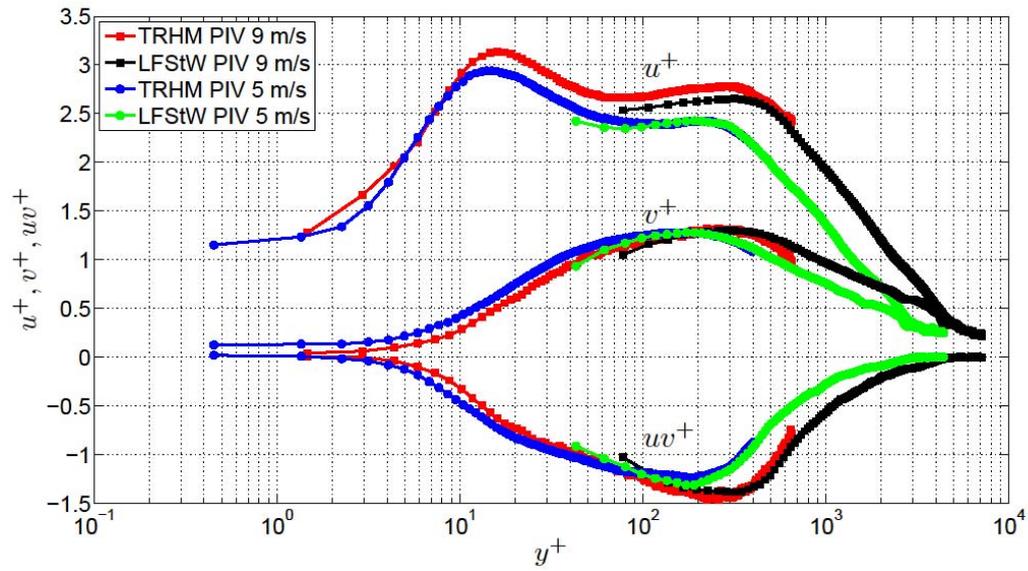

Figure 31: Streamwise and wall normal turbulence intensities and Reynolds shear stress profiles at s = 3.983 m in the APG region for both Reynolds numbers.

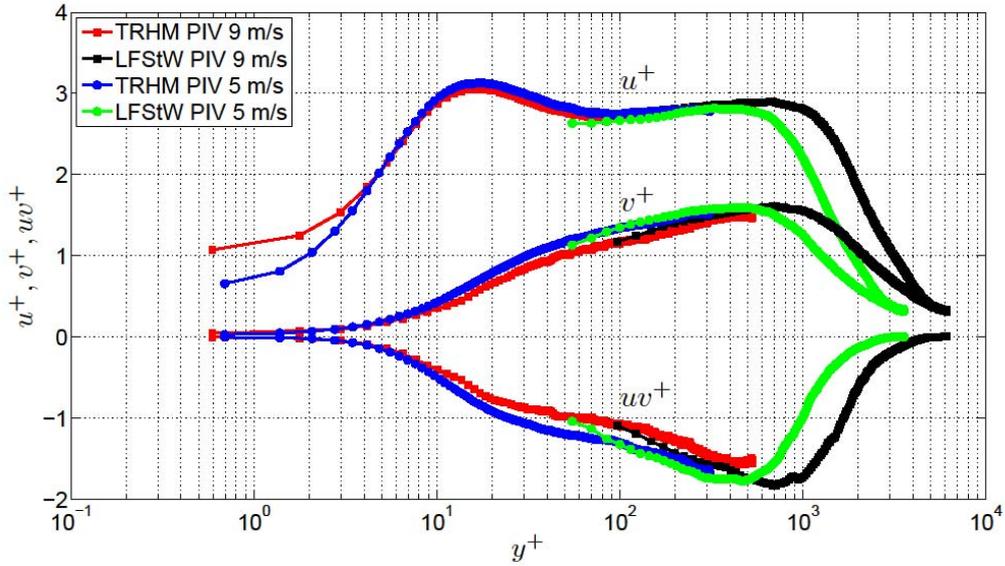

Figure 32: Streamwise and wall normal turbulence intensities and Reynolds shear stress profiles at s = 5.233 m in the APG region for both Reynolds numbers.

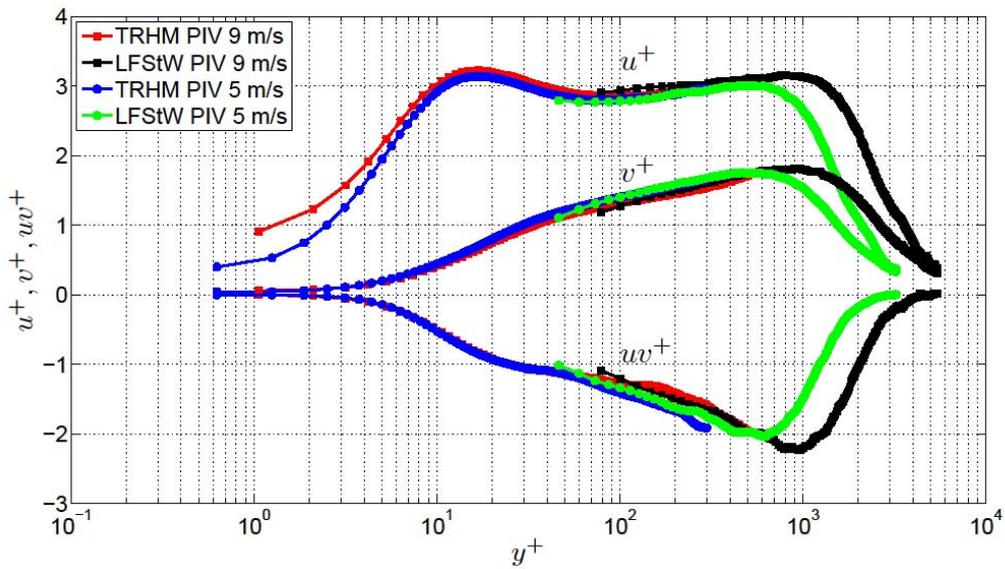

Figure 33: Streamwise and wall normal turbulence intensities and Reynolds shear stress profiles at s = 5.858 m in the APG region for both Reynolds numbers.

| S (m) | $U_\tau$ (m/s) at $U_\infty$ = 5 m/s | $U_\tau$ (m/s) at $U_\infty$ = 9 m/s |
|---|---|---|
| 3.983 | 0.26 | 0.46 |
| 5.233 | 0.21 | 0.36 |
| 5.858 | 0.19 | 0.32 |

Table 7 : Friction velocity measured with the high magnification PIV at the two Reynolds numbers.

Finally, spanwise stereo-PIV measurements were also carried out at two positions in the APG region (s = 4.095 and 5.692 m) with the set-up described in section 3.2. The measurements provide a field of view of 32.5 cm in the transverse direction and 12 cm in the wall-normal one. Here just the mean streamwise velocity fields are shown to analyze the transverse homogeneity. Figure 34 shows the evolution with z of this velocity component at both stations and for both Reynolds numbers. It should be firstly noted that the field of view is not big enough to capture the boundary layer thickness δ even at the first station s = 4.095 m. A trend similar to the one found for the inlet transverse field (section 9.2, Figure 42) is observed (very small variations of the free-stream velocity with z which give visible variations of δ). The trace of the turning vanes located in the plenum chamber is still influencing slightly the transverse homogeneity of the flow but not enough to consider the flow to be 3D. The Reynolds stresses (not shown) are not affected. A strong similarity is observed between the two Reynolds numbers.

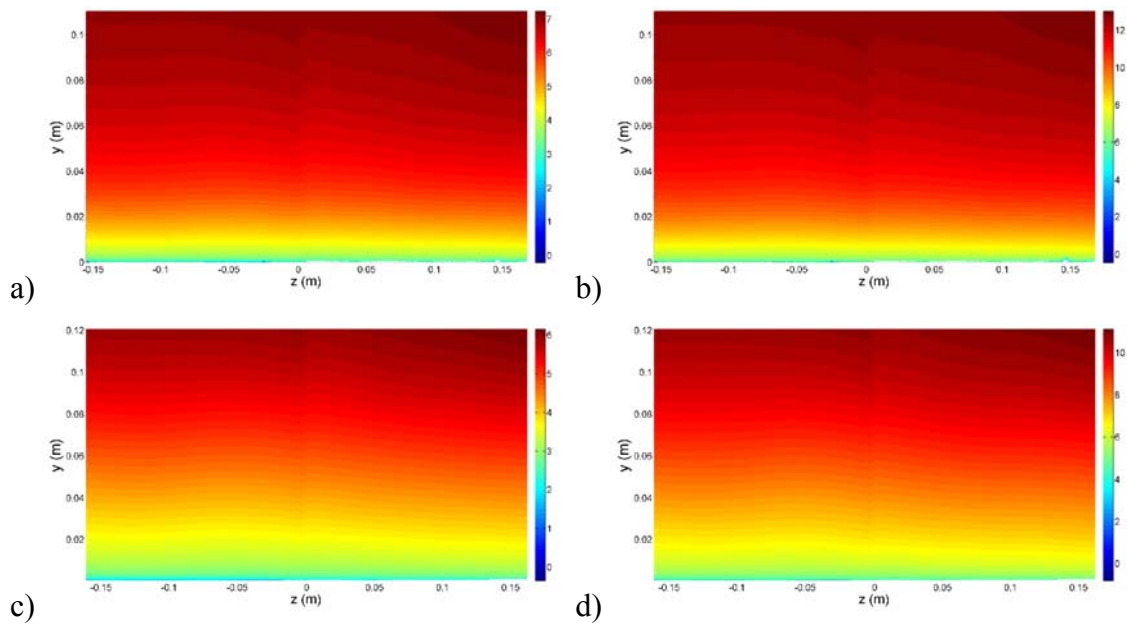

**Figure 34: Mean streamwise velocity fields in spanwise planes at two stations in the APG region and for both Reynolds numbers, a) s = 4.095, $U_\infty$ = 5 m/s, b) s = 4.095, $U_\infty$ = 9 m/s, c) s = 5.692, $U_\infty$ = 5 m/s and d) s = 5.692, $U_\infty$ = 9 m/s.**

## 7. Data synthesis

In order to provide a global view of the flow over the model, Figure 35 to Figure 39 combine the results of the FPG and APG experiments for the different measured quantities. To facilitate comparisons with numerical simulations, the velocities are given in the international units system (i.e. m/s) and in the wind tunnel reference frame (X: horizontal axis parallel to the wall of the wind tunnel, Y: wall normal axis). The field extends on about 5.5 m streamwise. Note that the Y axis is stretched for visibility purpose. It is interesting to remark that in the FPG region, despite the behaviour of the BL thickness evolution as evidenced in Table 3 and Table 4, the turbulence is developing slowly but continuously away from the wall. This development is of course much more significant in the APG region, but this time in relation with the mean velocity and BL thicknesses.

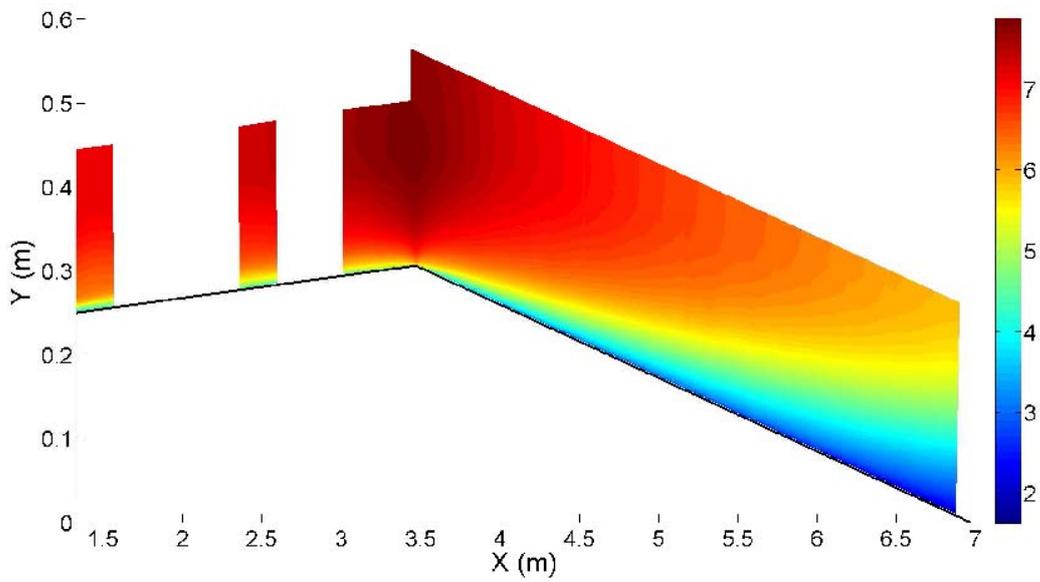

a)

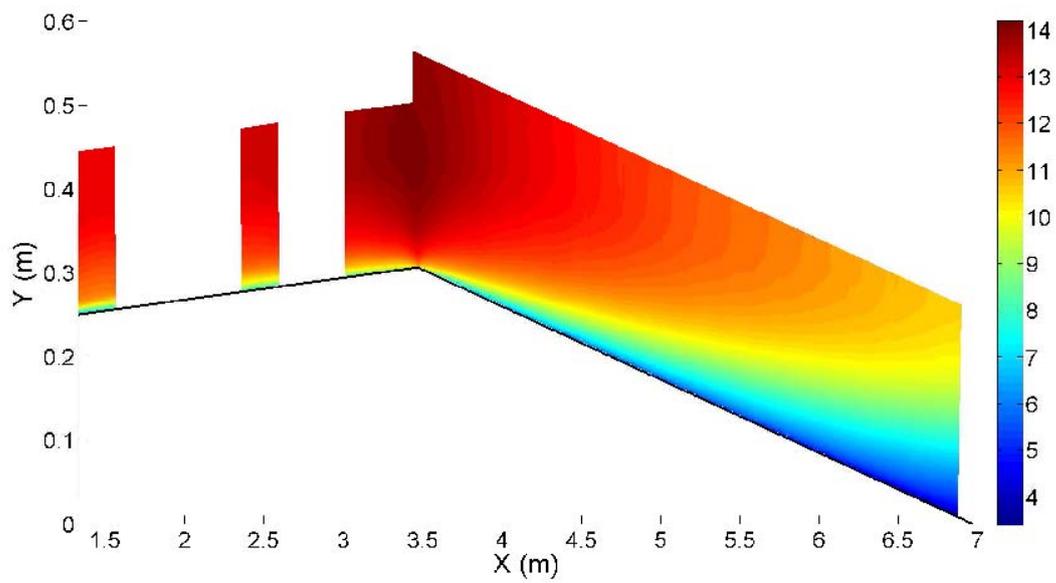

b)

**Figure 35: Mean horizontal velocity field above the ramp (in m/s), a) $U_\infty$ = 5 m/s, b) $U_\infty$ = 9 m/s.**

a)
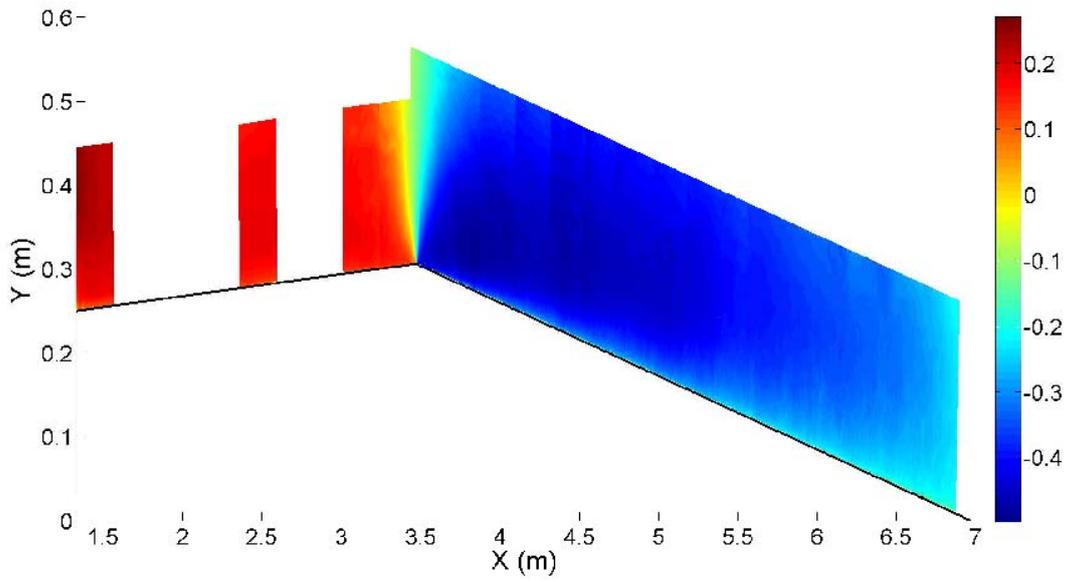

b)
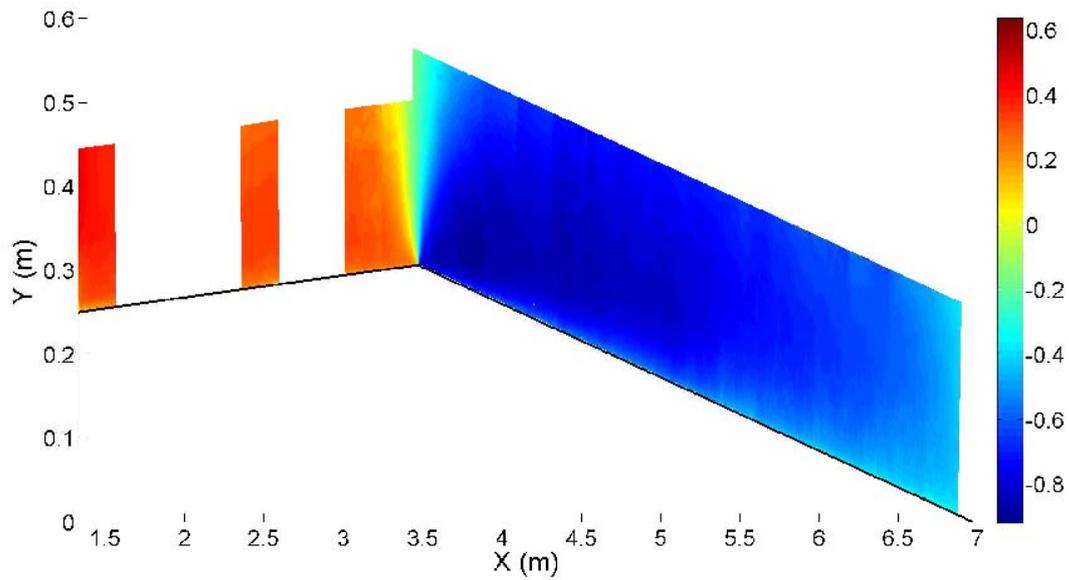

**Figure 36: Mean vertical velocity field above the ramp (in m/s), a) $U_\infty$ = 5 m/s, b) $U_\infty$ = 9 m/s.**

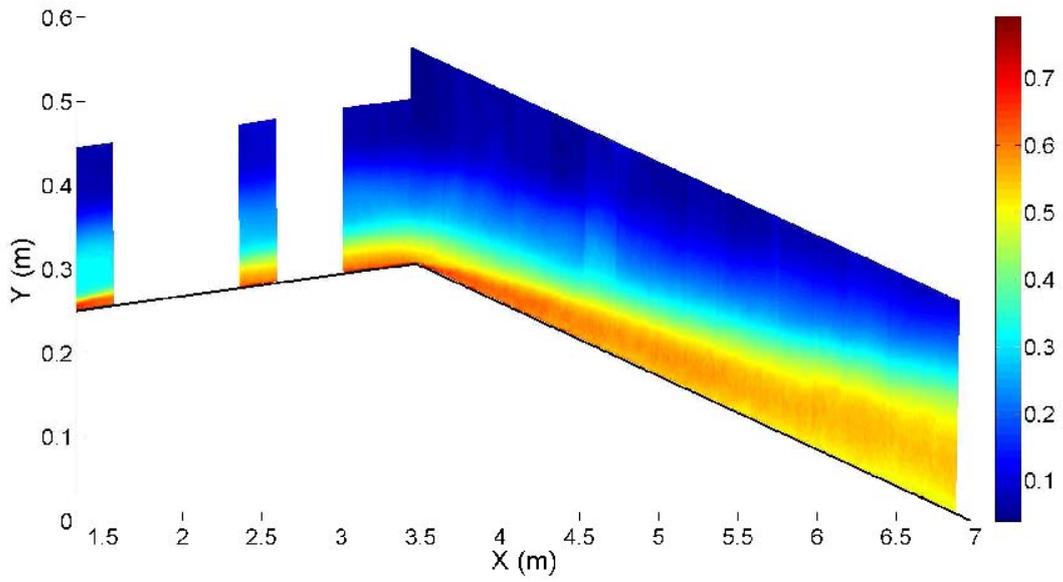

a)

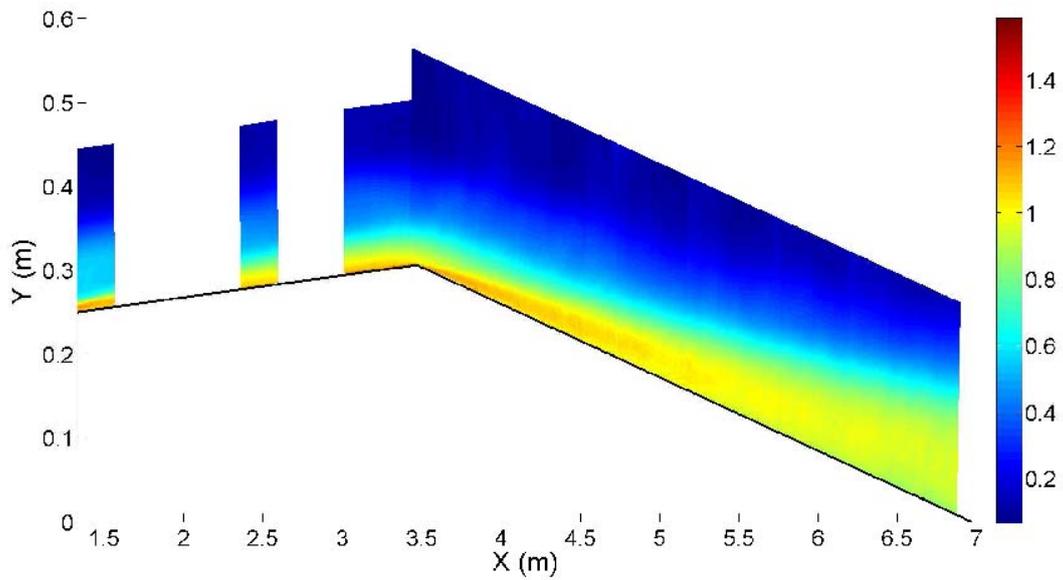

b)

**Figure 37: Horizontal turbulence intensity field above the ramp (in m/s), a) U$_\infty$ = 5 m/s, b) U$_\infty$ = 9 m/s.**

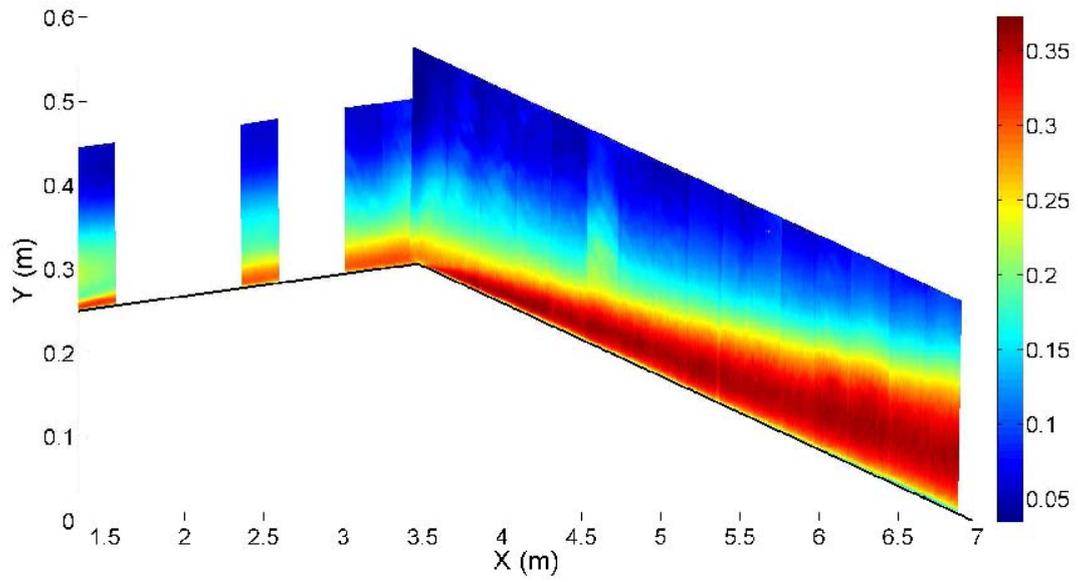

a)

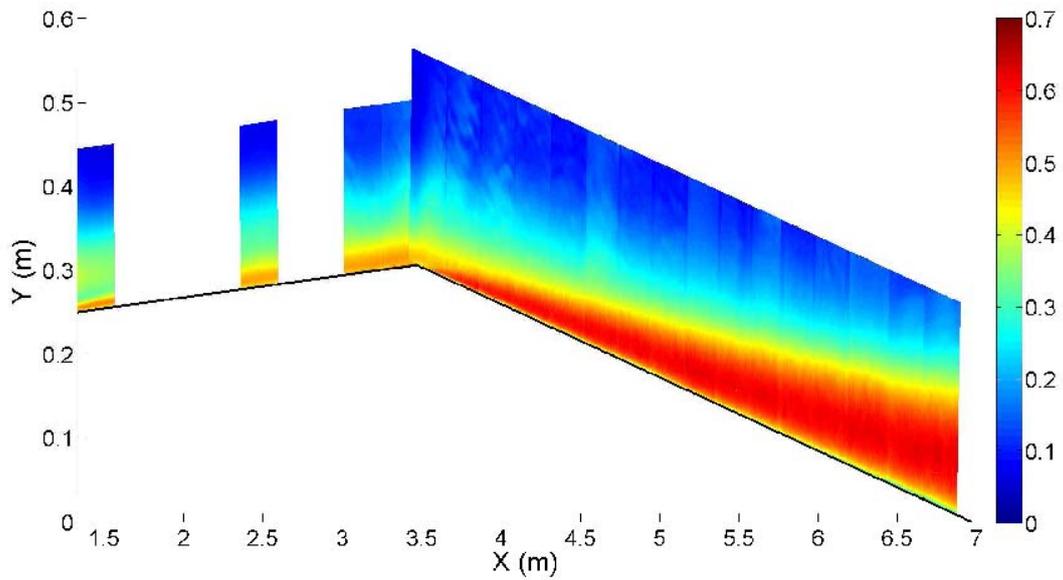

b)

**Figure 38: Vertical turbulence intensity field above the ramp (in m/s), a) $U_\infty$ = 5 m/s, b) $U_\infty$ = 9 m/s.**

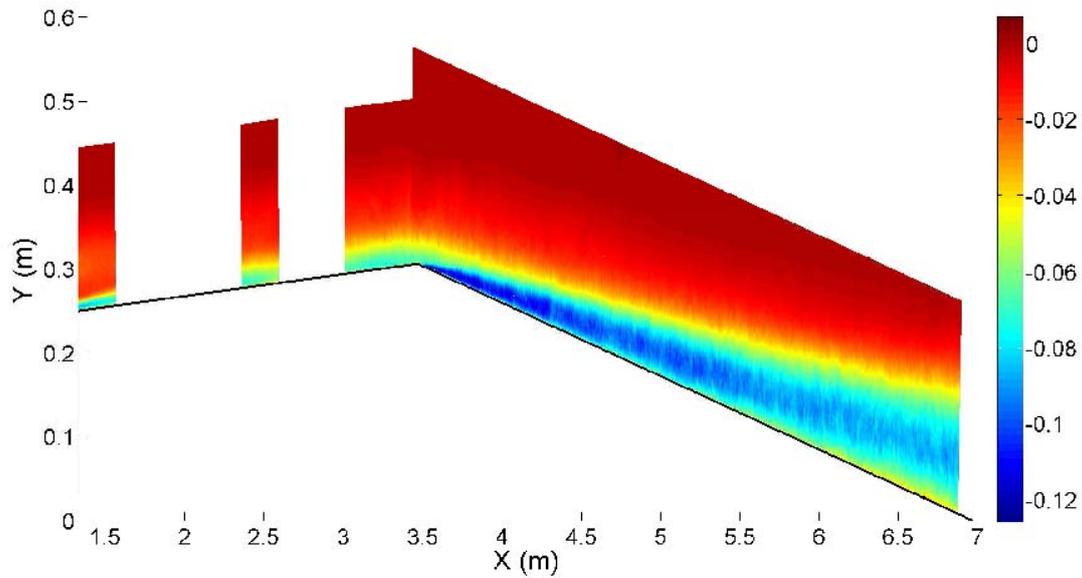

a)

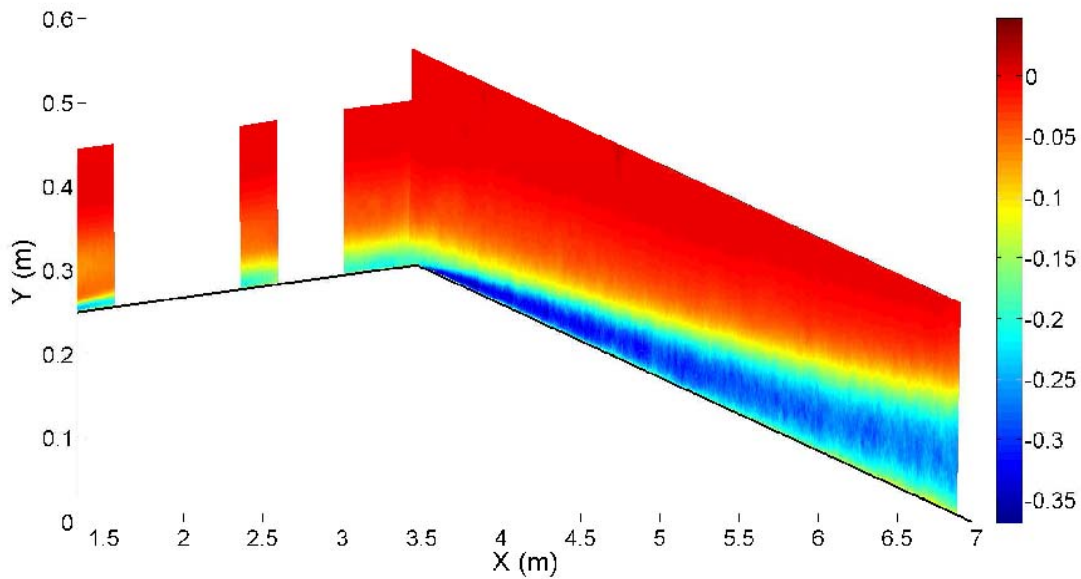

b)

**Figure 39:** $\overline{u'v'}$ **Reynolds shear stress field above the ramp (in m²/s²), a)** $U_\infty$ = 5 m/s, b) $U_\infty$ = 9 m/s.

## 8. Conclusion

As was emphasized in the introduction, near wall turbulence is clearly today the Achille heel of turbulence modelling. This is a flow region which is unavoidable in any industrial configuration and where turbulence models lack sufficient physical grounding to allow them to be really predictive. At the moment, they provide only week extrapolations around known situations. The EuHIT experiment was set by a consortium of researchers with two main objectives: -first to generate data allowing to better understand the physics of a decelerating boundary layer and - second to provide a challenging test case, representative of real conditions and allowing to improve turbulence models. The geometry of the model used is a simplified but representative mock-up of the flow over the suction side of an airfoil. A turbulent boundary layer is first accelerated and then decelerated without separation. The Reynolds numbers reached are high enough to be representative of real cases. The size of the studied boundary layer allowed detailed and accurate measurements inside it and down to very close to the wall. The first objective has been fulfilled: thanks to the original PIV set-up assembled by the consortium, 30 000 samples of high quality instantaneous velocity fields were recorded for two Reynolds numbers in a field that covers several BL thicknesses. The analysis of this unique data set is now under way by the partners in order to characterize the turbulence structures and to understand the underlying physics.

The present paper is aimed at fulfilling the second objective: providing a well characterized challenging test case to the turbulence modellers, both RANS and LES. For this purpose, several complementary measurements were performed to allow as complete as possible a characterization of the mean velocity and turbulence fields over the model. Besides, special care was taken to characterize at best the boundary conditions. The pressure distribution on both the model and the wind tunnel upper wall was characterized. The two-dimensionality of the flow around the mid vertical plane was checked both in terms of wall pressure, mean velocity and turbulence intensity. This two-dimensionality is quite good taking into account the size of the facility. The corner flow was characterized and does not show any structure able to perturb the flow around the mid vertical plane. The upstream boundary conditions on the lower wall were characterized in detail, providing reliable inlet conditions for RANS computations. The upper wall boundary layer is not characterized. Its tripping is very similar to the tripping of the lower wall so the bottom boundary conditions can be used with symmetry for the upper wall at the same station. As the pressure gradient encountered on the upper wall is much smoother than on the model, this should not have a significant influence on the result.

Looking at the data on the model itself, no measurements were performed in the curved converging part of the model due to the lack of optical access. This should not be an issue as this is not a region where turbulence models have a problem. All the challenging parts have been carefully characterized: - the FPG plate at some positions, including the very beginning, after the first strong variation of pressure gradient and the very end, just before the APG region. - The APG region, which was the main objective of the experiment, is characterized on its full length in terms of velocity and turbulence fields and at three stations very near the wall to assess the friction velocity. Over both regions, the turbulence can be considered out of equilibrium (and returning slowly toward it) with a physics which is far from the ZPG boundary layer physics which is used to build RANS (and near wall LES…) models. As expected and as evidenced already by several authors (Elsbery, et al., 2000) (Marquillie, et al., 2008), (George, et al., 2010a), (George, et al., 2010b), (Shah S.I., 2010), this APG flow evidences a second turbulence peak which has it origin in an instability developing after the change of sign of the pressure gradient. This second peak develops largely in the outer part of the BL as the flow develops downstream. It is based on a physics which is not in the RANS turbulence models at the moment. This turbulence peak is different from the one generated by a flow separation which is clearly due to a free shear layer instability. The present test case appears thus as very challenging in terms of turbulence modelling and the authors expect that it will contribute to improve the turbulence models used routinely in practical industrial applications.

In order to facilitate comparisons, all the results presented in this paper are fully available on the TURBASE database of the EuHIT European project. The following links should be used:

| Pressure distribution | https://turbase.cineca.it/turbase/default/#/view_dataset/43 |
| Inlet boundary conditions | https://turbase.cineca.it/turbase/default/#/view_dataset/42 |
| Near wall high mag. profiles | https://turbase.cineca.it/turbase/default/#/view_dataset/32 |
| Side wall corner flow | https://turbase.cineca.it/turbase/default/#/view_dataset/41 |
| FPG flow | https://turbase.cineca.it/turbase/default/#/view_dataset/44 |
| APG streamwise large field | https://turbase.cineca.it/turbase/default/#/view_dataset/25 |
| APG spanwise planes | https://turbase.cineca.it/turbase/default/#/view_dataset/45 |

**Acknowledgement**

The authors are thankful to the EuHIT EC project (http://www.EUHIT.org) for the financial support which has allowed this experiment to take place. The LML members of the team are thankful to CISIT, the *region Nord-Pas-de-Calais*, the European community and the CNRS for the financial support and notably the one which has allowed making the test section fully transparent, permitting this experiment. The Monash team gratefully acknowledges the support of an Australian Research Council support via a Discovery project grant in this research. J. Soria also gratefully acknowledges the support of an Australian Research Council Discovery Outstanding Researcher Award fellowship.

## 9. Annex: Characterization of boundary conditions
### 9.1. SPIV set-ups for upstream boundary conditions and side wall corner flow (UBC SPIV & CF SPIV).

In order to characterize upstream boundary conditions at 6.8 m from wind tunnel entrance, a stereo PIV experiment was carried out. The set-up corresponds to a spanwise wall-normal plane centred in the wind tunnel test section. The field of view was imaged by two Hamamatsu 2k by 2k cameras equipped with Nikon 105mm lenses at f# = 8 in forward scattering (see Figure 40). The light sheet, provided by a double pulse Innolas laser, was introduced into the wind tunnel through the opposite wall and was set tangent to the bottom surface to minimize reflections. Its thickness was about 1.5 mm.

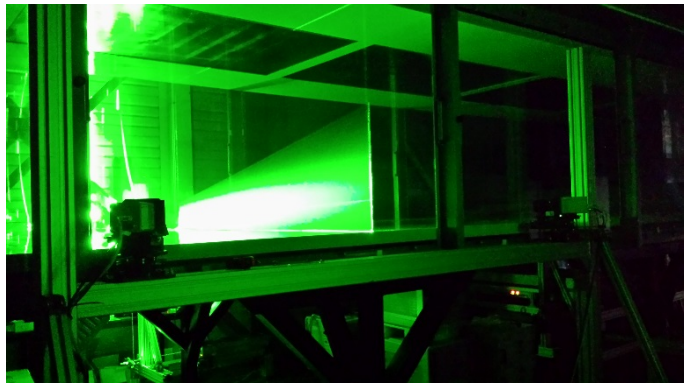

**Figure 40: Picture of the SPIV set-up used to obtain inlet boundary conditions at x = 6.8 m.**

The tests were performed for the two free stream velocities (5 m/s and 9 m/s at the wind tunnel entrance). For each velocity, 10 000 velocity fields were acquired to ensure good convergence of the mean flow and turbulence intensities. The data were processed by the modified version of the Matpiv toolbox by LML.

Also, in complement to the main experiments, a characterization of the corner flow which develops on the side walls of the wind tunnel was performed using a spanwise/wall-normal Stereo PIV plane of 235 mm height and 435 mm width. The streamwise positions of the investigated plane were on the model at s = 5595mm that is 2095 mm from the -5° ramp articulation (APG region) and upstream of the model at 6.8 m from tunnel entrance (flat plate region). Two Hamamatsu 2k by 2k cameras fitted with Nikon 50 mm lenses were used in Scheimpflug conditions and symmetric forward scattering. Figure 41 shows a picture of the set-up used. The light sheet, about 2.5 mm in thickness, was introduced in the wind tunnel through the opposite wall and was normal to the -5° ramp or the wind tunnel bottom wall respectively. It was tuned tangent to the wall surface to minimize reflections. The cameras aperture was $f_\# = 5.6$. Tests were performed for the two velocities of the experiment (5 m/s and 9 m/s at the wind tunnel entrance). For each velocity, 4000 velocity fields were acquired to ensure good convergence for the mean flow and turbulence intensities.

For both experiments the data were processed by the modified version of the Matpiv toolbox by LML. A self-calibration similar to the one proposed by (Wieneke, 2005) was applied with the Soloff reconstruction method (Soloff, et al., 1997).

For the upstream boundary conditions, the Multigrid/multipass cross-correlation PIV analysis ( (Willert & Gharib, 1991), (Soria, 1996)) was done with four passes starting with 48 x 64 pixels and ending with 16 x 24 pixels which was found to be the optimal final interrogation window size. Also, before the final pass, image deformation ( (Scarano, 2002), (Lecordier & Trinité, 2004)) was used to improve the quality of the results. The final interrogation window size corresponds to 2.4x2.4 mm² in the physical space. The mesh spacing was 1 mm in both directions corresponding to an overlap of about 60 %. This results in 180 points in the wall normal direction and 299 in the transverse one. The maximum displacement was about 11 pixels in the external region. The two light sheets were separated by about 0.6 mm to optimise the signal to noise ratio (Foucaut, et al., 2014).

For the corner flow, the Multigrid/multipass cross-correlation analysis ( (Willert & Gharib, 1991), (Soria, 1996)) was also done with four passes starting with 48 x 64 pixels and ending with 16 x 24 pixels which was found to be the optimal final interrogation window size. The final interrogation window corresponds to 3.5 mm² in the physical space. The mesh spacing was 1.5 mm in both directions corresponding to an overlap of about 60 %. This results in 156 points in the wall normal direction and 290 in the transverse one. A maximum displacement of 11 pixels was also chosen in the external region to ensure good results for the turbulence intensities.

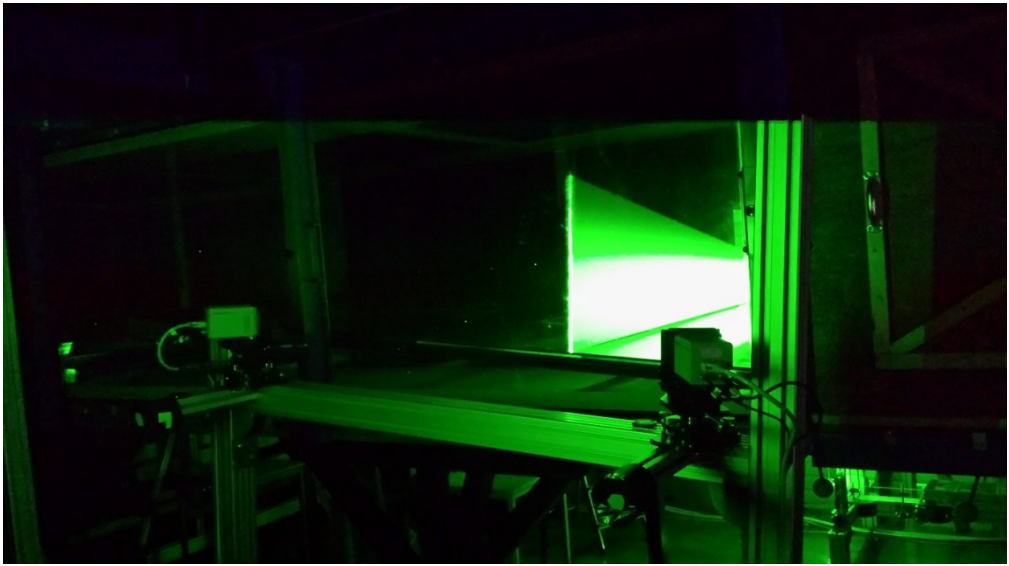

**Figure 41: Picture of the SPIV set-up used to characterize the corner flow upstream of the model and on the -5° ramp.**

## 9.2. Inlet boundary conditions at 6.8 m from wind tunnel entrance

In this section, results are presented in the (X,Y,Z) wind tunnel reference frame. Figure 42 shows the mean streamwise velocity field for both free-stream velocities investigated. The behaviour is the same. In both cases, a small variation in Z is observed. The local free-stream velocity is nearly constant (difference between minimum and maximum less than 0.3%), however the difference is more marked on the boundary layer thickness δ as the extraction of this quantity is very sensitive to small variations in the local free-stream velocity. At 5 m/s, the mean BL thickness is 10.7 cm with a minimum value of 9.9 cm and a max of 11.3 cm. At 9 m/s, the mean is 10.0 cm with a min of 9.4 cm and a max of 10.5 cm. These slight variations are attributed to the remains of the wakes of the turning vanes in the settling chamber.

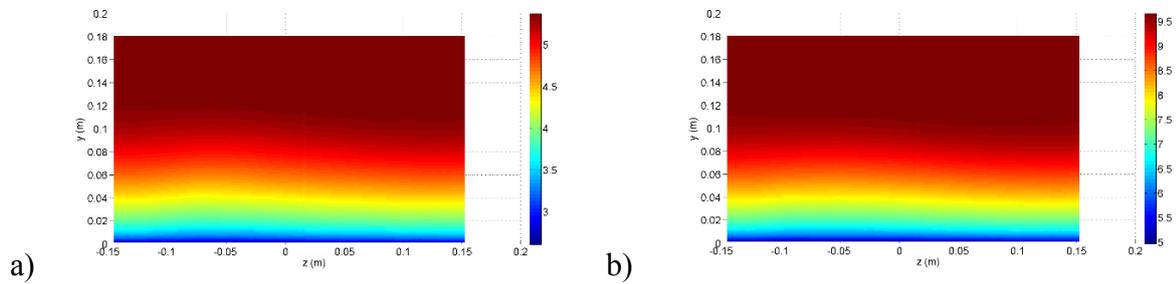

**Figure 42: Mean streamwise velocity field U in m/s for a) $U_\infty$ = 5 m/s and b) $U_\infty$ = 9 m/s in a spanwise-wall normal plane at X = 6.8 m (2.6 m upstream of the ramp).**

Figure 43 shows several mean velocity profiles for $U_\infty$ = 9 m/s. Similar curves are obtained for 5 m/s. The profile at Z = -0.065 m is plotted with error bars at ±0.5%, showing the good superposition of these profiles.

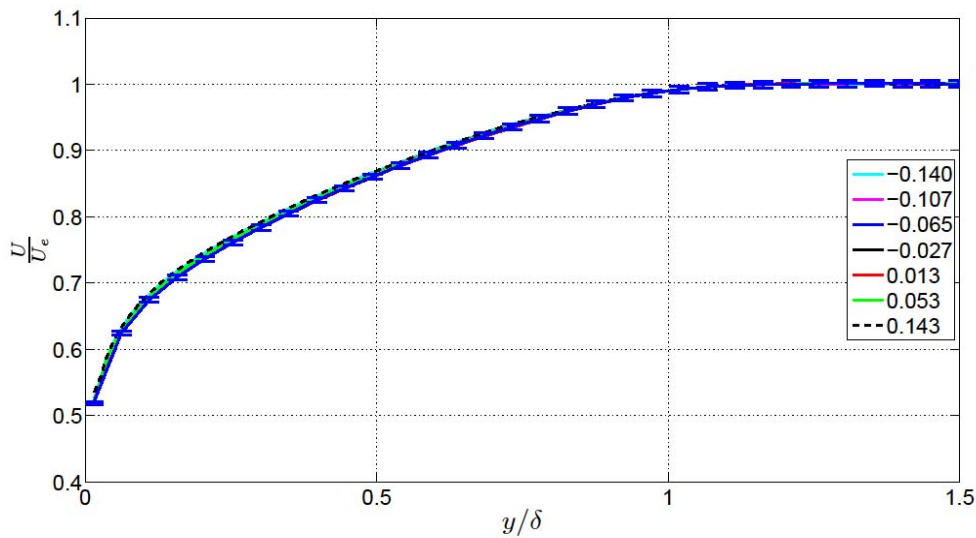

**Figure 43: Mean streamwise velocity profiles at different Z positions, at $U_\infty$ = 9 m/s and X = 6.8 m (2.6 m upstream of the ramp), scaled in outer units.**

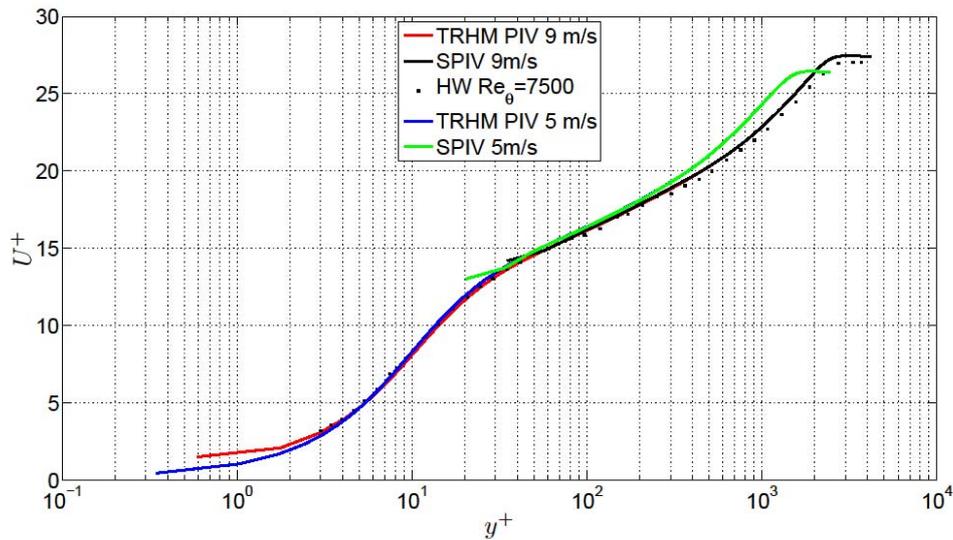

**Figure 44: Mean streamwise velocity profile (SPIV) at 2.6 m upstream of the ramp and at the center for both free stream velocities, completed with data obtained with the near wall set-up described in 3.3 (TRHM PIV), compared to hot-wire data (HW) at $Re_\theta$ = 7500 from (Carlier & Stanislas, 2005).**

Figure 44 shows for the two cases the streamwise velocity profile in the plane of symmetry, plotted in wall units and completed with the data obtained by TRHM PIV, compared to previous hot-wire data at $Re_\theta$ = 7500 (Carlier & Stanislas, 2005). A small adjustment of the friction velocity $u_\tau$ was necessary to connect smoothly the UBC SPIV and the TRHM PIV data: $u_\tau$ = 0.203 m/s was used for UBC SPIV compared to 0.204 m/s for TRHM PIV at 5 m/s and 0.3515 m/s and 0.350 m/s respectively at 9 m/s. This is explained by the fact that the two experiments were not conducted at the same time (about one month between both) which could change slightly the viscosity due to variations of atmospheric pressure. The collapse with the hot-wire data is good for the 9 m/s case as it corresponds to almost the same momentum Reynolds number (7750 compared to 7500). The boundary layer characteristics are then provided in Table 8.

| Reference velocity $U_\infty$ | $U_e$ (m/s) | $\delta$ (mm) | $\delta^*$ (mm) | $\theta$ (mm) | H | $Re_\theta$ | $u_\tau$ (m/s) | $\beta$ |
|---|---|---|---|---|---|---|---|---|
| 5 m/s | 5.36 | 109. | 18.0 | 13.1 | 1.37 | 4700 | 0.203 | -0.069 |
| 9 m/s | 9.64 | 102. | 16.4 | 12.0 | 1.37 | 7750 | 0.350 | -0.049 |

**Table 8: Inlet boundary layer characteristics**

The same analysis was also performed for the four main components of the Reynolds stress tensor, showing also good collapse between the UBC SPIV and the TRHM PIV data and reasonable agreement with the previous HW data at Re = 7500.

As a synthesis, the combined profiles of the streamwise mean velocity, the turbulent kinetic energy ($k = \frac{1}{2}\overline{u_i'^2}$) and the production ($\overline{u'v'}\frac{\partial \bar{U}}{\partial y}$) are given in wall units in Figure 45 to Figure 47 (The individual Reynolds stresses can be provided if needed). The friction velocity is given in Table 8. Supposing a local equilibrium, the dissipation can be approximated by the production term.

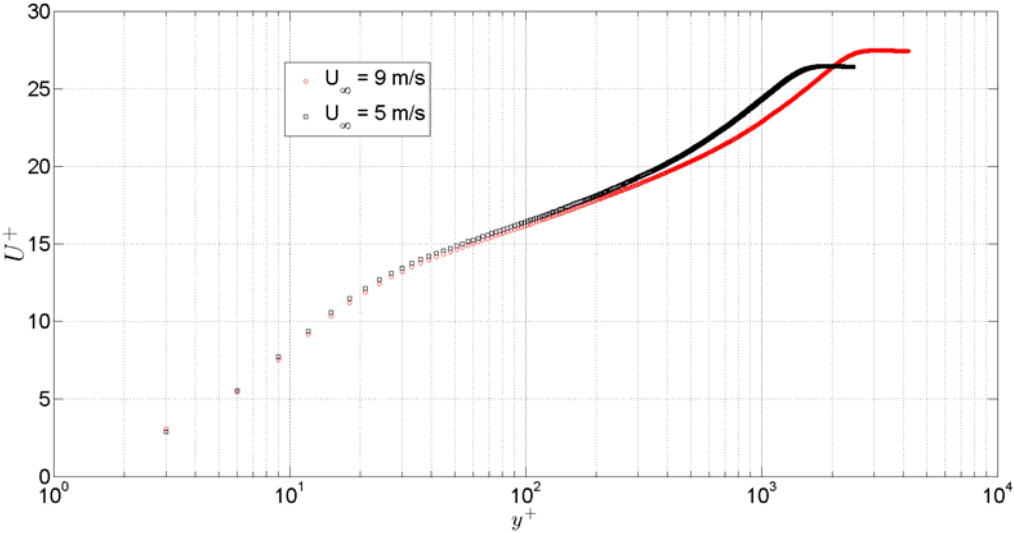

**Figure 45 : Mean streamwise velocity profile in wall-units for both velocities studied at 2.6 m upstream of the ramp**

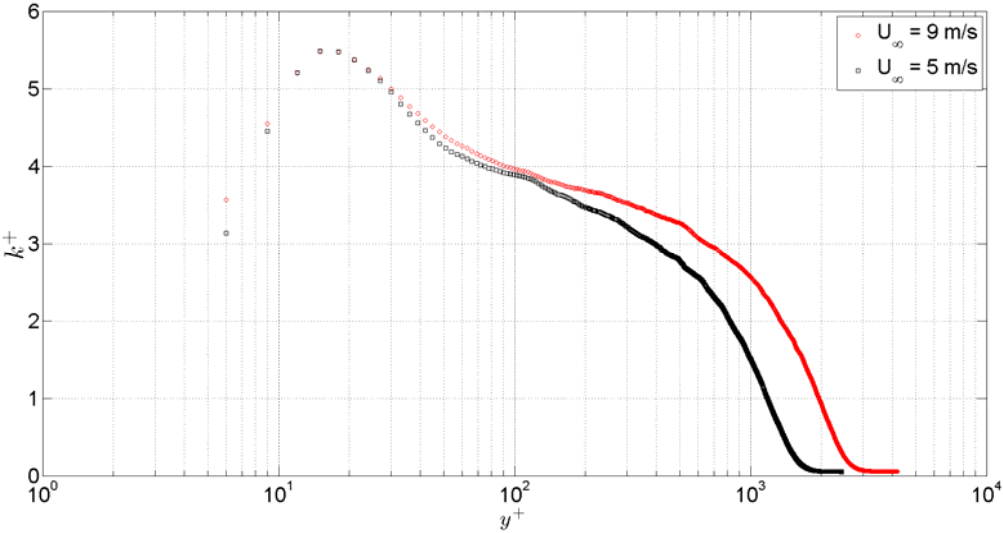

**Figure 46: Turbulent kinetic energy profile in wall-units for both velocities studied at 2.6 m upstream of the ramp**

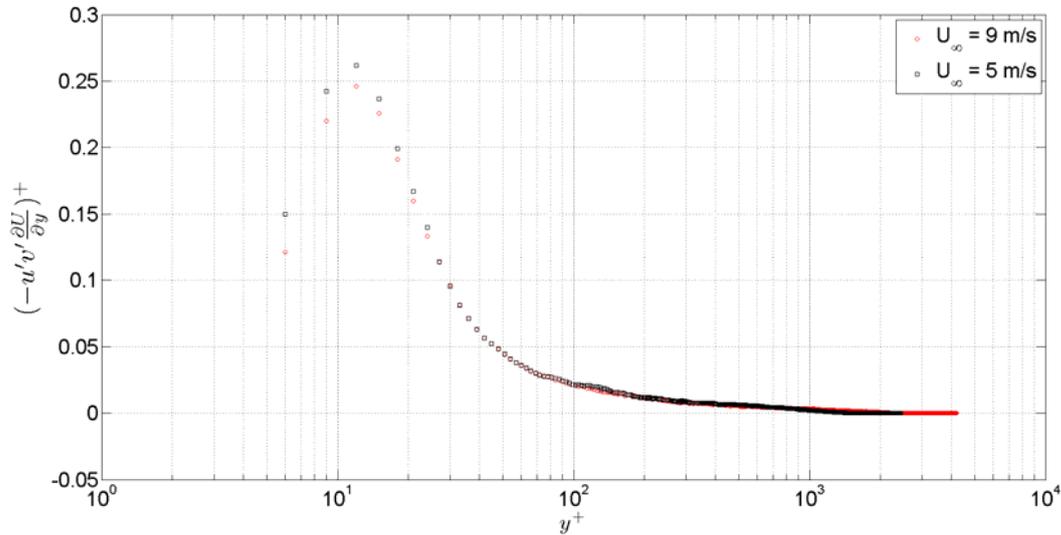

**Figure 47: Main production term profile in wall-units for both velocities studied at 2.6 m upstream of the ramp**

### 9.3. Characterization of the side wall corner flow.

For the result above the -5° ramp, local coordinates attached to the wall of the ramp are used. The x direction is the direction parallel to the ramp, y is the wall-normal direction and z the transverse one, with z = 0 corresponding to the wind tunnel centreline. The side wall investigated corresponds then to z = 1 m. For the result at X = 6.8 m, the (X,Y,Z) coordinate system of the wind tunnel is used. The side wall investigated is at Z = 1 m.

Figure 48 shows the mean streamwise velocity field U, the wall-normal (V) and spanwise (W) ones obtained near the side wall at $U_\infty$ = 9 m/s at 6.8 m from wind-tunnel. The free-stream velocity shows a slight variation from Z = 0.55 to Z = 0.8 m. Around Z = 0.7 m it is slightly smaller (less than 1% difference but visible). As the spacing obtained between the two "high speed extrema in the external regions" is about 200 mm, this variation is attributed to the turning vanes which are located inside the plenum chamber of the wind tunnel (10 blades distributed on the 2 m width). The results for $U_\infty$ = 5 m/s are not shown as they are similar to the 9 m/s case.

Figure 49**Erreur ! Source du renvoi introuvable.** shows the mean streamwise velocity field U, the wall-normal (V) and spanwise (W) ones obtained near the side wall at $U_\infty$ = 9 m/s on the -5° ramp at s = 5595 mm. The corner vortex is clearly evidenced in this figure. The centre is approximately at z = 0.93 m and y = 0.03 m. Beyond 200 mm from the side wall, the influence of the vortex is very small. This is confirmed by mean streamwise velocity U (color contour in Figure 49). The spanwise variation of the free-stream velocity shown at 6.8 m from tunnel entrance has almost disappeared. Same conclusions were obtained for the 5 m/s case which is not shown.

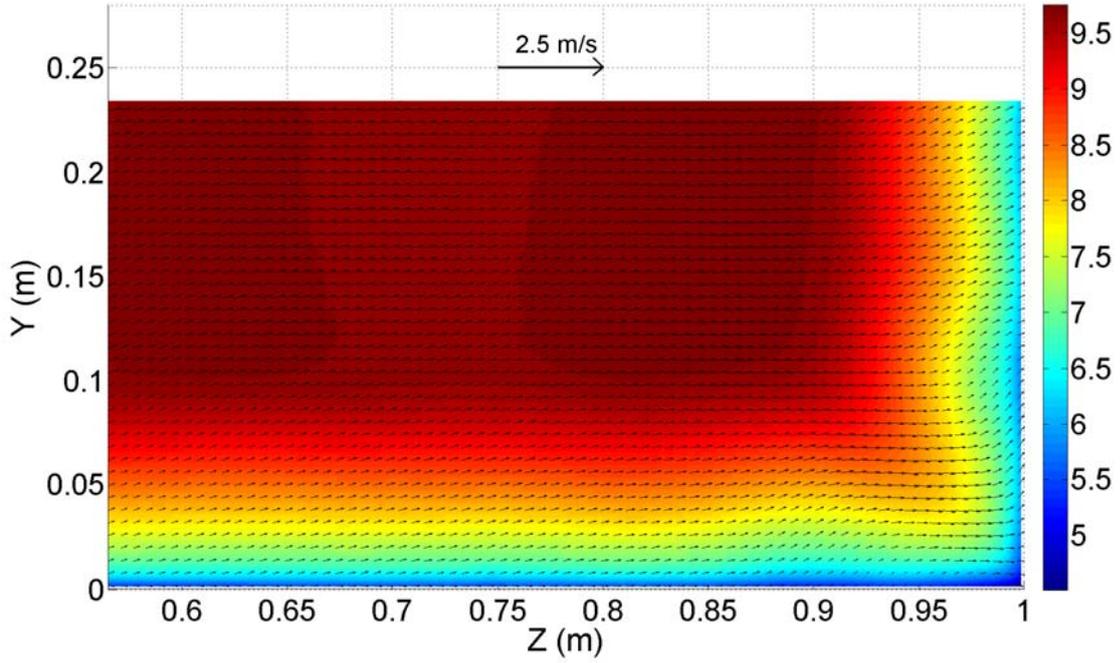

**Figure 48: Mean streamwise velociy U (color plot), wall-normal (V) and spanwise (W) ones (vectors plot) for $U_\infty$ = 9 m/s at 6.8 m from wind-tunnel entrance.**

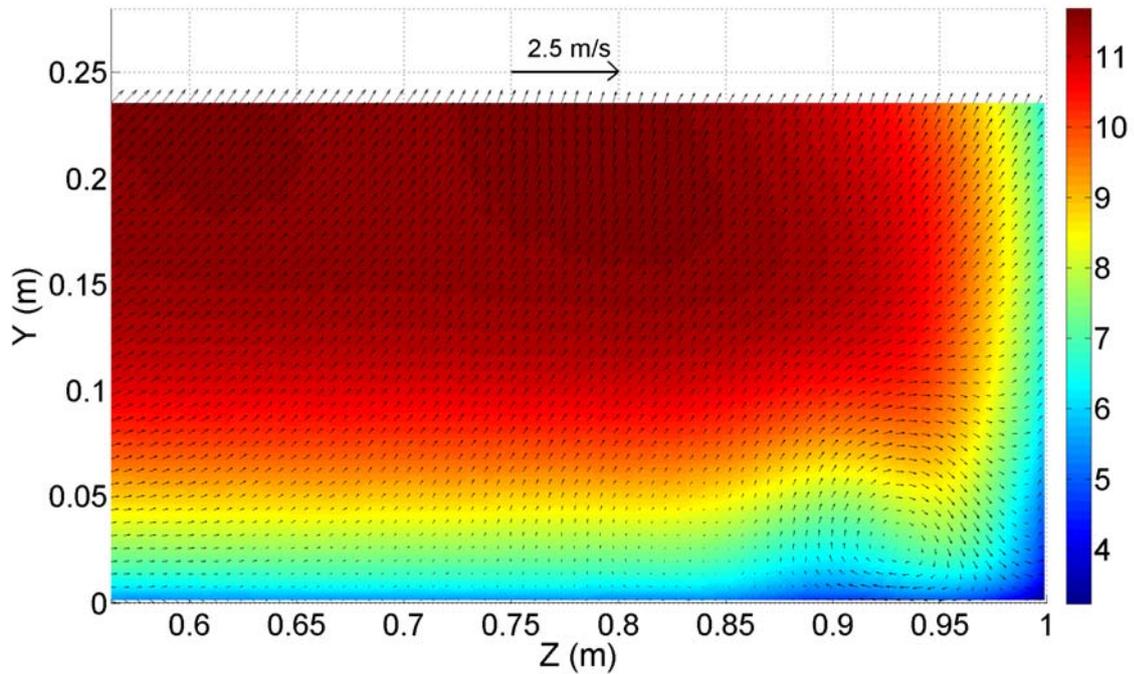

**Figure 49: Mean streamwise velociy U (color plot), wall-normal (V) and spanwise (W) ones (vectors plot) for $U_\infty$ = 9 m/s on the -5° ramp, s = 5595 mm.**